\documentclass[a4paper,11pt]{article}
\usepackage{jheppub} 

\usepackage{lineno}
\usepackage{braket}
\usepackage{booktabs}
\usepackage{xcolor}
\newcommand{\sun}[1]{{\color{purple}{#1}}}

\arxivnumber{} 

\title{\boldmath Multipartite entanglement characterizing topological phase transitions in holographic nodal line semimetals}

\author[a]{Xiantong Chen,}
\author[a,b]{Xuanting Ji,}
\author[a,c]{Ya-Wen Sun}
\affiliation[a]{
    School of Physical Sciences,\\ 
    University of Chinese Academy of Sciences, Beijing 100049, China
}
\affiliation[b]{
    Department of Applied Physics, College of Science, \\
    China Agricultural University, Beijing 100083, China
}
\affiliation[c]{
    Kavli Institute for Theoretical Sciences, \\
    University of Chinese Academy of Sciences, 
    Beijing 100049, China
}

\emailAdd{{chenxiantong23@mails.ucas.ac.cn}, {jixuanting@cau.edu.cn}, {yawen.sun@ucas.ac.cn}}

\abstract{
   Topological states of matter are characterized by nonlocal structures that are naturally encoded in the quantum entanglement of many-body wavefunctions. Topological semimetals are  short-range entangled states at weak coupling and their entanglement structure at strong coupling  remains largely unexplored. In this work, we investigate the multipartite entanglement structure of strongly coupled holographic nodal line semimetals. Building on previous studies of entanglement entropy and the holographic c-function, we focus on multipartite entanglement measures, including the conditional mutual information,  multi-entropy, and the Markov gap which is based on the entanglement wedge cross section. Our results demonstrate that while these multipartite measures vanish in the long-distance limit $l \to \infty$, which confirms that the holographic nodal line semimetal remains a short-range entangled state, their large $l$ scaling behavior remains highly sensitive to the underlying topology. The large $l$ power-law decay and scaling exponents serve as robust, non-local order parameters that exhibit sharp changes at the quantum critical point. This work establishes multi-partite entanglement as a powerful probe of quantum topological phase transitions in strongly coupled topological systems.
}

\begin{document}
\maketitle
\flushbottom

\section{Introduction}
    \label{Section1}
    The classification of topological states of matter represents a frontier in modern condensed matter physics, as these phases transcend the traditional Landau paradigm of spontaneous symmetry breaking \cite{ChernWeilTheoryzhang2001lectures}. Because topological phases lack a local order parameter, their characterization relies on global properties of the many-body wavefunction \cite{TopologicalPhaseMatterWitten:2015aoa}. Central to this distinction is the structure of quantum entanglement. Generally, topological matter is categorized into two classes: topologically ordered states, which exhibit long-range entanglement (LRE) \cite{Wen:2012hm, PhysRevLett.96.110405}, and symmetry-protected topological (SPT) phases \cite{2015ARCMP...6..299S}, which possess only short-range entanglement (SRE) but remain distinct from trivial states as long as specific symmetries are preserved \cite{TopologicalPhaseMatterWitten:2015aoa}.

Topological semimetals occupy a particularly subtle position in this landscape. Unlike gapped topological phases, they are gapless and host symmetry-protected band degeneracies \cite{TopologicalSemimetalswan}. Among them, nodal line semimetals are characterized by closed loops of degeneracy in momentum space, protected by symmetries such as the mirror symmetry \cite{PhysRevB.84.235126, TopologicalNodalLineSemimetalsFang_2016}. In weakly coupled settings, their low-energy physics admits a quasiparticle description in terms of Bloch Hamiltonians, and their topological properties can be encoded in band-theoretic invariants \cite{TopologicalNumberkm1,TopologicalNumberkm2,TopologicalNumbermoore}. From the real-space perspective, such systems are generally expected to belong to the class of short-range entangled states.

However, this picture becomes inadequate once strong interactions are introduced. In strongly correlated systems, quasiparticles may cease to exist and single-particle band topology is no longer a reliable organizing principle \cite{1999PhyB..259..353C, 2007NatPh...3..168B, Zaanenscience}. In this regime, entanglement is expected to play a central role in governing the low-energy physics, and the question of how topological properties are encoded in the many-body entanglement structure becomes both natural and pressing.

Holographic duality provides a powerful framework to address this question. By mapping a strongly coupled quantum field theory to a classical gravitational theory in a higher-dimensional spacetime, holography offers a controlled setting in which strongly correlated topological phases can be studied nonperturbatively \cite{HolographyMaldacena:1997re, GKPWgubser,GKPWwitten,HolographicCMTZaanen:2015oix}. Within this framework, holographic models of various kinds of topological semimetals, include Weyl semimetal \cite{HolographicWeylSemimetalsLandsteiner:2015lsa, HolographicWeylSemimetalsLandsteiner_2016,HolographicWeylSemimetalsLandsteiner:2016stv, HolographicWeylSemimetalsPlantz_2018}, nodal line semimetal \cite{Liu:2018bye, HolographicNodalLineSemimetalsLiu_2021}, Weyl-$\boldsymbol{Z}_2$ semimetal \cite{Ji:2021aan}, Weyl-Nodal line coexisting semimetals \cite{HolographicWeylNodalChu_2024} have been constructed and phase diagrams for quantum topological phase transitions have been obtained \cite{Landsteiner:2019kxb}. In holography, the entanglement entropy of a boundary region is geometrized by the area of a minimal Ryu-Takayanagi (RT) surface in the bulk \cite{RyuTakayanagiFormulaRyu_2006}, and it has been proven to be a powerful way to capture both of the physical properties of gravity \cite{Almheiri_2021} and many-body systems \cite{Albash:2012pd, Jeong:2022jmp, Yang:2025qkh, Yang:2025upg}. While for the holographic semimetals, previous studies have successfully utilized the $c$-function, a measure derived from the entanglement entropy of a strip, as a diagnostic for quantum topological phase transitions \cite{Baggioli:2020cld,CFunctionOrderParameterBaggioli_2023}. 

In strongly coupled systems, correlations are expected to be highly collective, suggesting that multipartite entanglement may play a more fundamental role. While the $c$-function provides a coarse-grained view of the degrees of freedom along the renormalization group (RG) flow \cite{DifferentialEntropyLiu_2014}, it does not fully reveal the rich multipartite entanglement that define strongly correlated topological matter.  Moreover, large-scale entanglement behavior is closely tied to the infrared structure of the theory and may encode universal information about criticality and phase transitions that is invisible to short-distance probes. This long range behavior is also important for us to detect if strongly coupled topological semimetal states remain short range entangled states. 

Motivated by these considerations, the primary goal of this work is to systematically investigate the multipartite entanglement structure of holographic nodal line semimetals. We focus in particular on tripartite entanglement measures and study how their behavior evolves across the quantum topological phase transition. In this work, we employ three complementary classes of multipartite entanglement measures. First, we study the conditional mutual information (CMI), a tripartite measure that quantifies the correlation between two disjoint regions conditioned on a third system. The conditional mutual information has been widely used in condensed matter systems to distinguish short-range entangled states from those with long-range entanglement. It has also played an important role in studying the entanglement structures in holography \cite{ MultiEntanglementUpperBoundju2024holographicmultipartiteentanglementupper, ConditionalMutualInformationJu:2024kuc} and in connecting the boundary entanglement behavior with bulk geometry \cite{DifferentialEntropyMyers_2012, ju2024generalizedrindlerwedgeholographic,DifferentialEntropyBalasubramanian_2014, ConditionalMutualInformationJu_2024, ConditionalMutualInformationJi_2025}. Second, we analyze the holographic multi-entropy, which generalizes entanglement entropy to multiple disjoint regions and captures higher-order correlation structures. We specially focus on one tripartite measure $\kappa$, which is built from the multi-entropy, and isolates genuine tripartite entanglement among the three parties composing a pure state \cite{GenuineMulipartiteiizuka2025genuinemultientropy,GenuineMulipartiteiizuka2025genuinemultientropyholography}. Third, we consider the entanglement wedge cross section (EWCS), which is dual to the reflected entropy and entanglement of purification on the boundary \cite{ReflectedEntropydutta2019canonicalpurificationentanglementwedge, Umemoto2018}. We also consider one special multipartite entanglement measure derived from EWCS: the Markov gap, which detects tripartite entanglement structures that are not locally unitary to triangle states \cite{MarkovGapHayden_2021}. These tripartite measures are all finite quantities without UV divergences, whose behavior does not depend on the choice of UV regularization \cite{MarkovGapju2025holographicmultipartiteentanglementstructures}.

Our primary objective is to examine the behavior of these measures across the topological phase diagram, with a specific focus on their scaling behavior at large distance scales $l$. In the limit where $l \to \infty$, we observe that all investigated tripartite measures vanish. This result confirms that the holographic NLSM remains a short-range entangled state, consistent with its identification as an SPT phase rather than an intrinsically topologically ordered LRE state. Crucially, however, we find that the asymptotic scaling behavior of these measures at large $l$ provides a sensitive probe of the system’s IR physics. The rate at which these multipartite correlations decay reflects the underlying topological phase and the emergence of the scaling behavior in the IR regime. Therefore, these measures serve as robust non-local order parameters in addition to the topological invariants \cite{HolographicTopologicalInvariantLiu_2018, Chen:2025akz, Chen:2025ahb} for topological phase transitions in the absence of a quasiparticle description. Finally, the holographic NLSM provides a suitable framework for observing the holographic UV-to-IR flow of multipartite entanglement, offering new perspectives on how entanglement is redistributed as the system evolves from a high-energy UV boundary to a topologically non-trivial IR fixed point.

The remainder of this paper is organized as follows. In Section \ref{Section2}, we review the holographic nodal line semimetal model and the computation of entanglement entropy and the c-function. Section \ref{Section3} is devoted to the conditional mutual information and its scaling behavior at large $l$. In Section \ref{Section4}, we study tripartite entanglement using the multi-entropy based measure $\kappa$ and analyze its large $l$ scaling behavior. Section \ref{Section5} focuses on the entanglement wedge cross section and one particular measure  derived from it: the Markov gap. We conclude in Section \ref{Section6} with a discussion and open directions for future work.

\section{Review of the holographic nodal line semimetal and the c-function}
    \label{Section2}
    In this section, we review the holographic framework for nodal line semimetals \cite{HolographicNodalLineSemimetalsLiu_2021} and the entanglement-based c-function that can be used to characterize their quantum topological phase transitions \cite{CFunctionOrderParameterBaggioli_2023}. We begin by briefly introducing the holographic model of nodal line semimetals and the associated bulk geometry, which captures the essential features of the topological and trivial phases as well as the critical regime separating them. We then summarize the computation of holographic entanglement entropy using the Ryu–Takayanagi prescription, focusing on the case of an infinite strip on the boundary, for which translational symmetry renders the extremal surface problem analytically tractable. Based on this setup, we review the definition and evaluation of the entanglement c-function, which measures the scale dependence of entanglement across the strip. As previously shown, this c-function exhibits sharp and characteristic behavior across the quantum topological phase transition and can be naturally interpreted as an entanglement-based order parameter for the transition. This section sets the stage for our subsequent analysis of multipartite entanglement structures in holographic nodal line semimetals.
\subsection{The  holographic topological nodal line semimetal}
    The bulk action for the holographic  nodal line semimetal is \cite{HolographicTopologicalInvariantLiu_2018}
        \begin{equation}
            \label{NodalHolographicModel}
            \begin{aligned}
                S_{\text{nodal}}&=\int d^5x\sqrt{-\det g}\left[\frac{1}{2\kappa^2}\left(R+\frac{12}{L^2}-\frac{{F_V}^2}{4}-\frac{{F_A}^2}{4}\right)\right.\\
                &+\left.\frac{\alpha}{3}\epsilon^{abcde}A_a\left(3{F}_{Vbc}{F}_{Vde}+{F}_{Abc}{F}_{Ade}\right)-(D_a\Phi)^* D^a\Phi-{m}^2|\Phi|^2-\frac{\lambda}{2}|\Phi|^4\right.\\
                &-\left.{m_B}^2 {B_{ab}}^*B^{ab}-\lambda_B|\Phi|^2{B_{ab}}^*B^{ab}-\frac{1}{6\eta}\epsilon^{abcde}\left(iB_{ab}{H_{cde}}^*-i{B_{ab}}^*H_{cde}\right) \right],
            \end{aligned}
        \end{equation}
        where $\kappa^2$ is the 5-dimensional gravitational constant, $L$ is the AdS radius.  $g$ is the metric for 5-dimensional AdS spacetime and $R$ is the corresponding Ricci curvature. $B$ is a complex 2-form field and $H_{abc}=\partial_a B_{bc}+\partial_b B_{ca}+\partial_c B_{ab}-iq_B{A}_aB_{bc}-iq_B{A}_bB_{ca}-iq_B{A}_cB_{ab}$. 
        $m_B$ and $\lambda_B$ are the coefficients of the potential for $B$. $\eta$ is a coupling constant. $V$ is a vector gauge field and $A$ is an axial gauge field with $F_V,F_A$ their corresponding field strengths. The vector gauge field $V$ and the axial gauge field $A$ do not play a role in the holographic nodal line semimetal, so without loss of generality they are set to zero here. $\alpha$ is the coupling constant for Chern-Simons term. $\Phi$ is a scalar field providing the effective mass term for the holographic nodal line semimetal. $m$ and $\lambda$ are the coefficients of the potential for $\Phi$. $D_a=\partial_a-iqA_a$ is the covariant derivative for $\Phi$ and $q$ is the corresponding coupling constant. 

        Without loss of generality we set  $2\kappa^2=L=1$, and fix the coupling constants for the Chern-Simons terms $\alpha=1,~\eta=2$, the mass term $m^2=-3, m_B=1$ and the coupling constants for the scalar fields $\lambda_B=1,~\lambda=\frac{1}{10}$, the coupling constants for axial gauge fields $q=1,~q_B=1$.
        
        {Before proceeding, we would like to introduce how to build the holographic model \eqref{NodalHolographicModel}. The holographic model is motivated by the field theory Lagrangian for topological nodal line semimetals}
        \begin{equation}
            \mathcal{L}=\bar{\Psi}\bigl(\gamma^\mu\partial_\mu -m -e\gamma^\mu V_\mu+\gamma^5\gamma^\mu A_\mu-\gamma^{\mu\nu}b_{\mu\nu}+\gamma^{\mu\nu}\gamma^5b_{\mu\nu}^5\bigr)\Psi,
        \end{equation}
        {where $b_{\mu\nu}$ is an antisymmetric real tensor, and the term $\gamma^{\mu\nu}b_{\mu\nu}$ contributes to the formation of the nodal line semimetal, $M$ is the mass term and $\Psi$ is a four component spinor.  According to the duality relation between the anti-symmetric tensor $\Bar{\Psi}\gamma^{\mu\nu}\gamma^5\Psi=-\frac{i}{2}{\epsilon_{\alpha\beta}}^{\mu\nu}\Bar{\Psi}\Gamma^{\alpha\beta}\Psi$ in the four-dimensional Minkowski spacetime, a pure imaginary dual part of the $b_{\mu\nu}$ needs to be introduced, which is ${b_{\mu\nu}^5}$.}

        The bulk field $B$ employed in \eqref{NodalHolographicModel} is dual to the fermionic bilinear $\bar\Psi \gamma^{\mu\nu}b_{\mu\nu} \Psi$, which is responsible for producing the topologically nontrivial nodal ring in the system and contribute the divergenvce for the axial current $J^\mu_5$
        \begin{equation}
             \partial_\mu J^\mu_5 = -2m \bar\Psi \gamma^5 \Psi - 2 b_{\mu\nu} \bar\Psi \gamma^{\mu\nu}\gamma^5 \Psi.
        \end{equation}
        {The coupling term between $B$ and $\Phi$ is determined by dimensional analysis and the operator product expansion (OPE) on the boundary \cite{Alvares:2011wb} \footnote{{\cite{Alvares:2011wb} provides a solid basis for the utilisation of the dual form field and its associated action when studying the dual operator \(\bar{\psi}\Gamma^{\mu\nu}\psi\) and \(\bar{\psi}\Gamma^{\mu\nu}\Gamma^5\psi\) in the context and AdS/QCD. Here we utilize the same two form field as we need the same bilinear operator, though our work focuses on a fundamentally different problem: strongly correlated topological semimetals.}}. Note that the $B$ field does not couple to the vector gauge field $V_{\mu}$ as the operator dual to $B$ is a charge neutral one while it carries an axial charge so that it still needs to be coupled to the axial gauge field $A_\mu$.}
    
        {The model \eqref{NodalHolographicModel} is sufficiently general to treat both holographic Weyl semimetals and holographic nodal line semimetals within a unified framework. This paper focuses primarily on holographic nodal line semimetals, so all terms related to Weyl semimetals, such as $A$ and $V$, are set to zero, and the corresponding Chern–Simons term does not affect the dynamics of the system. After introducing the motivation for the holographic model \eqref{NodalHolographicModel}, we will present its boundary behavior and various phases, based on the following ansatz for the holographic nodal line semimetal}

        \begin{equation}
            \label{BackGroundZeroTemperatureAnsatz}
            \begin{aligned}
                ds^2&=-u dt^2+f (dx^2+dy^2)+udz^2+\frac{dr^2}{u},\\
                B&=\frac{1}{2}\left(B_{xy}dx\wedge dy+iB_{tz}dt\wedge dz\right),~\Phi=\phi,
            \end{aligned}
        \end{equation}
        where all the nonzero components $u,f,B_{xy},B_{tz},\phi$ are real functions of the radial coordinate $r$.
        
        Under the ansatz \eqref{BackGroundZeroTemperatureAnsatz}, the boundary asymptotic behaviors of the fields are explicitly given below        \begin{equation}
            \label{BackGroundBoundaryBehaviour}
            \begin{aligned}
                \lim_{r\to\infty} \frac{u}{r^2}&=\lim_{r\to\infty} \frac{f}{r^2}=1,\\
                \lim_{r\to\infty}\frac{B_{xy}}{r}&=\lim_{r\to\infty}\frac{B_{tz}}{r}=b,~\lim_{r\to\infty}r\phi=M.
            \end{aligned}
        \end{equation}
       
        The dimensionless quantity $M/b$ is an important dimensionless order parameter that characterises the phase of the nodal line semimetal. The phase structure for the specific parameters that we pick above is as follows: the system gives a topologically non-trivial nodal line semimetal state at $M/b<0.8597$, is a critical point at $M/b=0.8597$, and gives a topologically trivial semimetal state at $M/b>0.8597$. Different phases are governed by the different  IR behaviors in the corresponding bulk  geometries of the holographic model \eqref{NodalHolographicModel}. Using the ansatz \eqref{BackGroundZeroTemperatureAnsatz}, the IR geometry for the topologically non-trivial phase is \cite{HolographicNodalLineSemimetalsLiu_2021}
        \begin{equation}
            \label{NonTrivialInfraredGeometry}
            \begin{aligned}
                u&=\frac{11+3\sqrt{13}}{8}~r^2(1+\delta u~ r^{\alpha_1}),\\
                f&=\sqrt{\frac{2\sqrt{13}}{3}-2}~r^{\alpha}(1+\delta f~ r^{\alpha_1}),\\
                B_{tz}&=\frac{\sqrt{54+15\sqrt{13}}}{8}~r^2(1+\delta b_{tz}~ r^{\alpha_1}),\\
                B_{xy}&=r^{\alpha}(1+\delta b_{xy}~ r^{\alpha_1}),\\
                \phi&=\phi_{0}~ r^{\beta},
            \end{aligned}
        \end{equation}
        where $\alpha=0.183,~\alpha_1=1.273,~\beta=0.228$ and $(\delta f,~\delta b_{tz},~\delta b_{xy})=(-2.616,~1.720,~-0.302)\delta u$. $\delta u$ and $\phi_{0}$ are shooting parameters. 
        We typically set $\delta u=-1$ and adjust $\phi_{0}$ to generate a family of solutions. Using the IR geometry as the boundary condition for the near horizon region, we numerically integrate the equations of motion for the bulk fields $u,f,B_{tz},B_{xy},\phi$ in the holographic model \eqref{NodalHolographicModel} to obtain their complete profiles. From the UV asymptotics of these solutions \eqref{BackGroundBoundaryBehaviour}, we extract the corresponding value of $M/b$. A key finding from the numerical calculations is that, for this specific IR geometry (the non-trivial phase), the value of $M/b$ never exceeds the critical value of 0.8597. This upper bound is in perfect agreement with the theoretical understanding that the IR geometry fundamentally determines the phase of the system, an understanding which itself characterizes the phase as topologically non-trivial.
        
        Similiarly, the IR geometry for the critical phase is
        \begin{equation}
            \label{CriticalInfraredGeometry}
            \begin{aligned}
                u&=u_c ~r^2(1+\delta u ~ r^{\beta}),\\
                f&=f_c ~r^{\alpha_c}(1+\delta f ~ r^{\beta}),\\
                B_{tz}&=b_{tzc}~r^2(1+\delta b_{tz} ~ r^{\beta}),\\
                B_{xy}&=b_{xyc}~r^{\alpha_c}(1+\delta b_{xy} ~ r^{\beta}),\\
                \phi&=\phi_c(1+\delta\phi ~ r^\beta),
            \end{aligned}
        \end{equation}
        where $\alpha_c=0.314,~\beta=1.274,~u_c=2.735,~f_c=0.754,~b_{tzc}=1.437,~b_{xyc}=1,~\phi_c=0.557$ and $(\delta u,~\delta f,~\delta b_{tz},~\delta b_{xy})=(0.882,~-2.151,~1.718,~-0.254)\delta\phi$. Without loss of generality, we set $\delta \phi=-1$. Applying a similar numerical protocol—solving the full bulk equations with this new critical IR geometry as the boundary condition—we extract the $M/b$ and confirm that it indeed yields the critical value of 0.8597 \cite{HolographicNodalLineSemimetalsLiu_2021}. This IR geometry is a Lifshitz-type solution, and its inherent scale invariance holographically corresponds to the scale-invariant behavior exhibited by the system at the quantum critical point.

        At last, the IR geometry for the topological trivial phase is
        \begin{equation}
            \label{TrivialInfraredGeometry}
            \begin{aligned}
                u&=\left(1+\frac{3}{8\lambda}\right)r^2,\\
                f&=r^2,\\
                B_{tz}&=\left(1+\frac{3}{8\lambda}\right)b_0~r^{2\sqrt{2}\frac{3\lambda_B+\lambda}{\sqrt{\lambda(3+8\lambda)}}},\\
                B_{xy}&=b_0~r^{2\sqrt{2}\frac{3\lambda_B+\lambda}{\sqrt{\lambda(3+8\lambda)}}},\\
                \phi&=\sqrt{\frac{3}{\lambda}}+\phi_1~r^{2\left(\sqrt{\frac{3+20\lambda}{3+8\lambda}}-1\right)},
            \end{aligned}
        \end{equation}
        where $\lambda,\lambda_B$ are defined in \eqref{NodalHolographicModel} as coupling parameters and $b_0, \phi_1$ are shooting parameters. Generally we set $b_0=1$ and vary $\phi_1$ to generate profiles for the bulk fields $u,f,B_{tz},B_{xy},\phi$, whose $M/b$ are never less than the critical value of 0.8597 as expected.
        \begin{figure}[htbp]
            \begin{minipage}{0.49\linewidth}
                \centering
                \includegraphics[width=\textwidth]{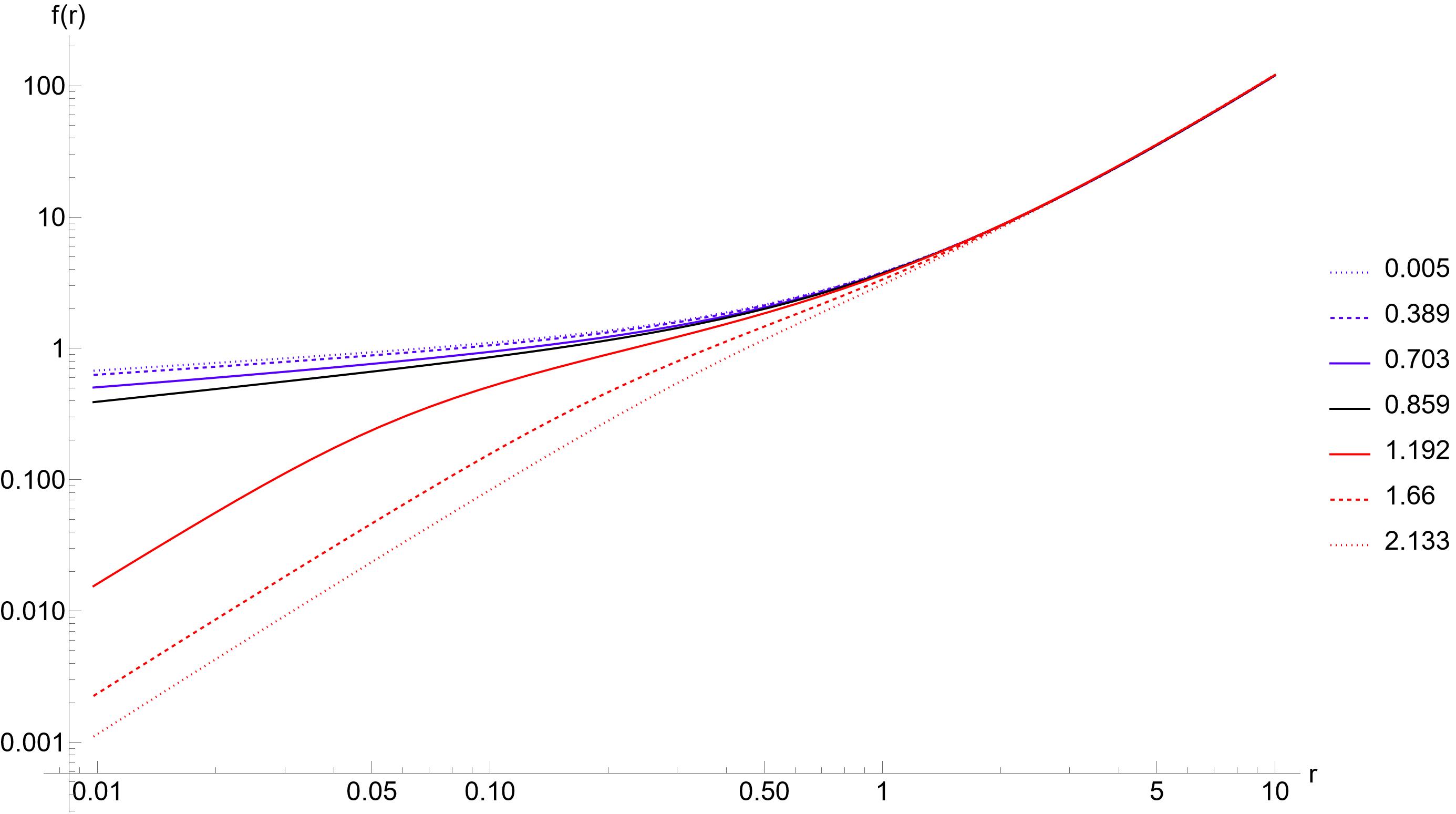}
            \end{minipage}
            \begin{minipage}{0.49\linewidth}
                \centering
                \includegraphics[width=\textwidth]{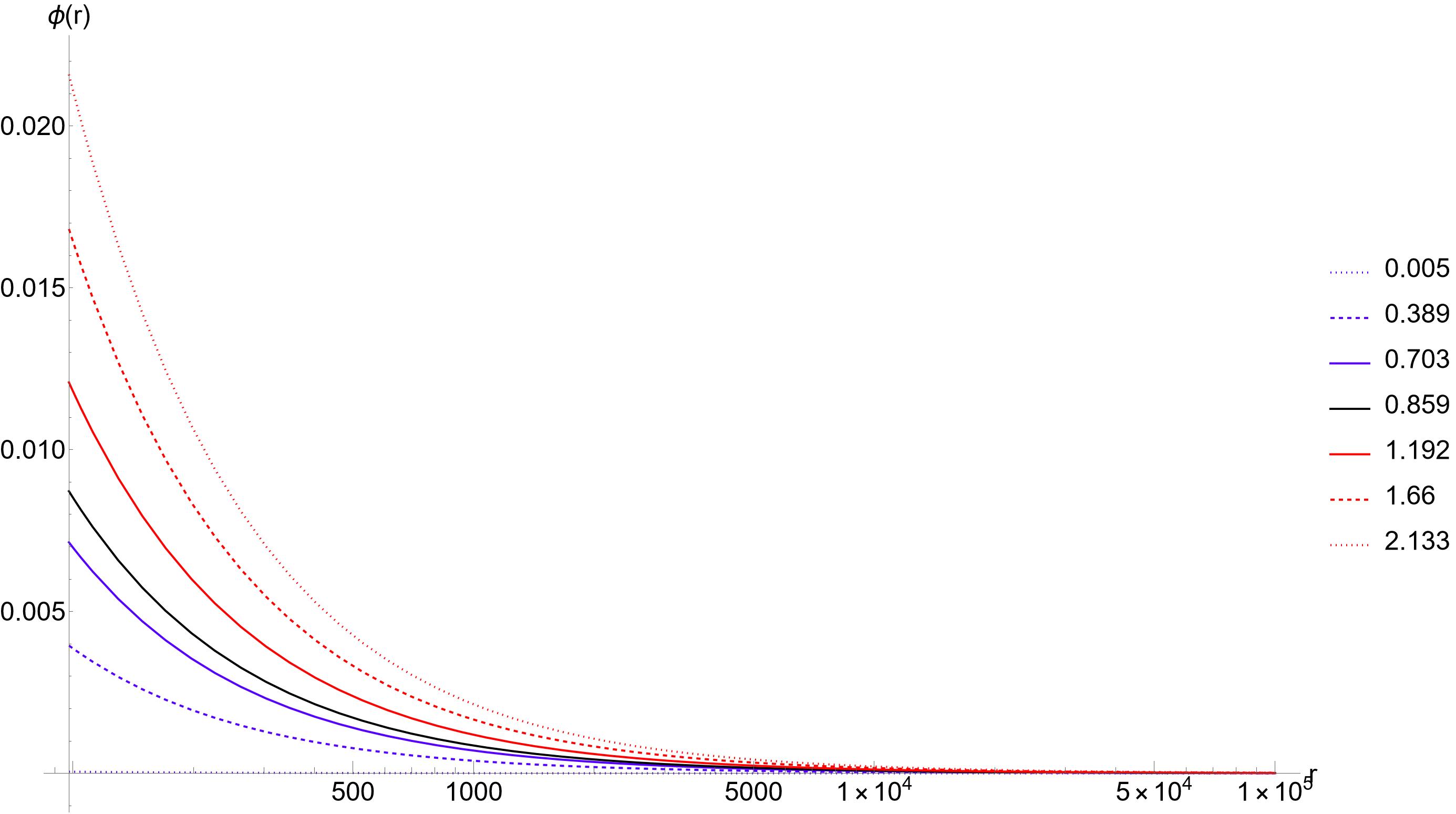}
            \end{minipage}
            \caption{Left: The profile of $f(r)$ for different values of $M/b$. Right: The profile of $\phi(r)$ for different values of $M/b$. {The profile for critical phase is shown in black, corresponding $M/b$ is 0.8597.}}
            \label{FigureBulkGeometryProfile}
        \end{figure}
        
        The radial coordinate $r$ in the holographic bulk naturally encodes the energy scale of the renormalization group (RG) flow, where the evolution from the UV boundary ($r=\infty$) to the IR horizon corresponds to the coarse-grained transition from high-energy degrees of freedom to low-energy macroscopic phases. In the vicinity of the topological phase transition, the bulk profiles exhibit a characteristic behavior dictated by the underlying quantum criticality. Specifically, as the control parameter $M/b$ approaches its critical value, the bulk solution's trajectory remains in close proximity to the critical fixed-point solution over an increasingly extended range of the radial scale.

        Due to the very continuous nature of the quantum topological phase transition, the deviation of the near-critical solutions from the exact critical profile occurs progressively deeper in the IR when $M/b$ approaches the critical value. This behavior signifies that the system ``lingers" near the scale-invariant critical phase before eventually flowing toward its respective IR destination (either the topological or trivial phase). This behavior, as illustrated in Fig.\,\ref{FigureBulkGeometryProfile}, provides a clear geometric manifestation of the RG flows of various phases.
        
        The low energy, large scale behavior of the system is governed by the IR geometry. Consequently, the large scale scaling behavior is determined by the scaling behavior of the IR geometry. Under a scale transformation, the IR geometry exhibits anisotropic scaling symmetry. Specifically, when we perform the transformation $r^{-1}\to s^{\mathbf{z}} r^{-1}$, the IR geometry transforms as $(t,z)\to s^{\mathbf{z}}(t,z)$ in the $t$ and $z$ directions, and as $(x,y)\to s(x,y)$ in the $xy$ directions. The scaling exponent $\mathbf{z}$ measures the anisotropy of the system's ground state and can be read off directly from the explicit form of the IR geometry. Its values for the topologically nontrivial, critical, and topologically trivial phases are
        \begin{equation}
            \label{InfraRedScale}
            \mathbf{z}=\{\frac{2}{\alpha},~\frac{2}{\alpha_c},1\}=\{10.929,~ 6.3694,~ 1\},
        \end{equation}
        respectively, where $\alpha$ is defined in \eqref{NonTrivialInfraredGeometry} and $\alpha_c$ is defined in \eqref{CriticalInfraredGeometry}.
    \subsection{The RT surface and the holographic c-function}
        Building upon the holographic nodal-line semimetal model introduced in equation \eqref{NodalHolographicModel}, this subsection reviews the calculation of entanglement entropy via the Ryu-Takayanagi (RT) prescription and the definition of the associated holographic c-function. To render the extremal surface problem tractable, we consider an infinite strip of finite width on the boundary. This choice is standard in holographic analyses for two key reasons. First, due to the translational symmetry along the strip, the corresponding co-dimension-2 RT surface in the 4+1-dimensional bulk becomes effectively one-dimensional, making its profile completely integrable. Second, the entanglement entropy of such a strip is a well-studied observable in boundary field theories, capturing non-local correlations at a scale set by its width.
        
        We work on a canonical Cauchy slice with the metric $g_{xx} dx^2+g_{yy} dy^2+g_{zz} dz^2+g_{rr} dr^2$, and denote the coordinates as $(x,y,z,r)=(x^1,x^2,x^3,r)$. Without loss of generality we consider a strip of width $l_i$, aligned along the $x^i$ direction, defined by the interval $x_i\in \left[-\frac{l_i}{2},\frac{l_i}{2}\right]$. Owing to the simple topology and translational symmetry of the bulk geometry, the corresponding RT surface can be parameterised by the boundary coordinates $x^j$ as $(x^1,x^2,x^3,r)=(x^i,x^2,x^3,r(x^i))$. The induced metric for the RT surface is then given by
        \begin{equation}
            \label{ExtremalSurfaceInducedMetric}
            ds^2=\sum_{j}g_{jj}(r)dx^jdx^j+g_{rr}(r)\left(\frac{dr}{d x^i}\right)^2 dx^idx^i.
        \end{equation}
        
        Since the induced metric \eqref{ExtremalSurfaceInducedMetric} does not depend on the boundary coordinate $x^j(j\ne i)$, the action for the RT surface can be reduced to a one-dimensional functional 
        \begin{equation}
            \label{ExtremalSurfaceAction}
            A=L^2\int_{-\frac{l_i}{2}}^{\frac{l_i}{2}}\sqrt{\prod_{j\ne i}g_{jj}\left(g_{rr}\left(\frac{dr}{d x^i}\right)^2+g_{ii}\right)} dx^i,
        \end{equation}
        where $L$ is the cutoff length for the transverse coordinates $x^j(j\ne i)$. The translational symmetry along $x^i$ implies that the corresponding Lagrangian density is independent of $x^i$. According to Noether's theorem, this leads to a conserved constant $C_i$. Denoting the integration in \eqref{ExtremalSurfaceAction} as the Lagrangian $\mathcal{L}$, we compute the conserved quantity conjugate to $x^i$
        \begin{equation}
            \label{ExtremalSurfaceConstant}
            C_i=\mathcal{L}-\frac{\partial\mathcal{L}}{\partial r'}r'=g_{ii}\sqrt{\frac{\prod_{j\ne i}g_{jj}}{g_{ii}+g_{rr}r'^2}},
        \end{equation}
        where $r'$ denotes $\frac{d r}{dx^i}$. {Given the strip on the boundary field theory, the conserved quantity $C_i$ is a constant on corresponding RT surface, hence $C_i$ depends only on the strip width $l_i$.} This reduced one-dimensional action is endowed with a constant of motion, making it fully integrable. The equation for the RT surface can thus be solved by analytic integration.
        
        The turning point $r_*$, defined by $r'(r_*) =0$, plays a vital role as the maximum depth the RT surface can approach in our computation. To ensure that the RT surface probes the near-horizon IR geometry, we require $r_*$ to be sufficiently close to the horizon. Consequently, we employ a reverse logic: instead of specifying the boundary width $l_i$ first, we begin by choosing a deep turning point $r_*$. The corresponding conserved quantity is then fixed at $C_i=\sqrt{\prod_i g_{ii}(r_*)}$. The boundary strip width $l_i$ that yields this specific RT surface is subsequently determined by integrate from $r_*$ to the boundary
        \begin{equation}
            \label{StripWidth}
            l_i=2\int_{r_*}^\infty \sqrt{\frac{g_{rr}{C_i}^2}{g_{ii}(\prod_{j=1}^n g_{jj}-{C_i}^2)}}dr.
        \end{equation}
        
\begin{figure}[htbp]
            \centering
            \includegraphics[width=0.6\textwidth]{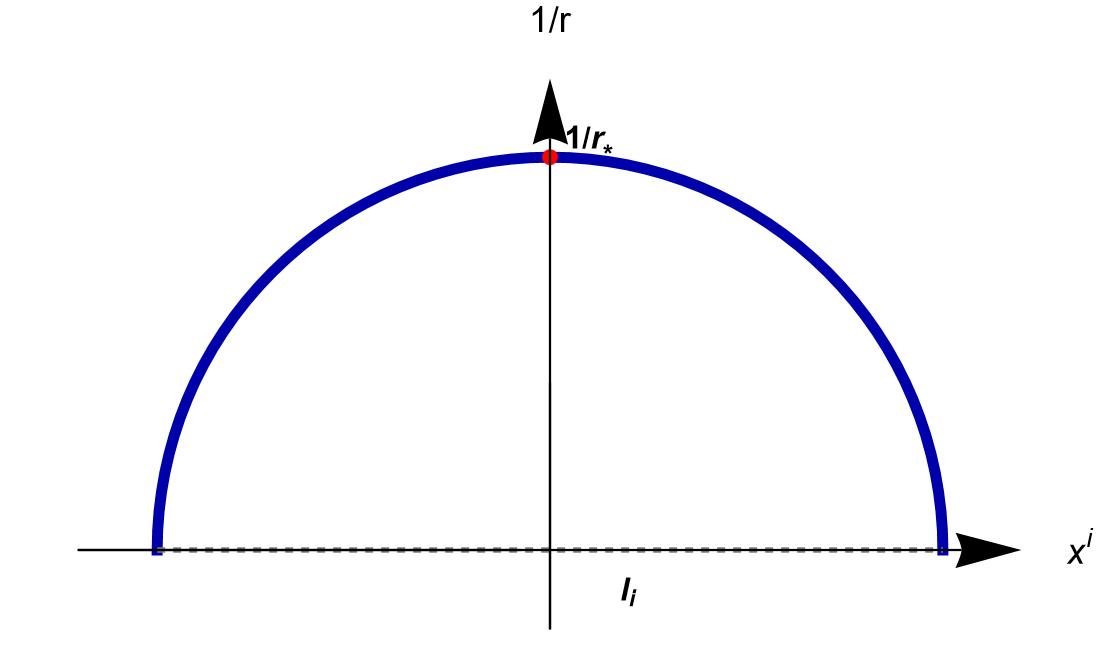}
            \caption{Illustration of an extremal surface anchored to a boundary strip along the $x^i$-direction. The surface is constructed by first choosing a turning point $r_*$ deep in the bulk; its corresponding boundary width $l_i$ is then derived from Eq. \eqref{StripWidth}.}
            \label{FigureEntanglementEntropy}
        \end{figure}

        Using the RT formula \cite{RyuTakayanagiFormulaRyu_2006}, the corresponding entanglement entropy is given by
        \begin{equation}
            \label{StripEntanglementEntropy}
            S_i=\frac{2L^2}{4G}\int_{r_*}^\infty\prod_{j=1}^n g_{jj}\sqrt{\frac{g_{rr}}{g_{ii}(\prod_{j=1}^n g_{jj}-{C_i}^2)}}dr,
        \end{equation}
        where $G$ is the gravitational constant. In summary, the algorithm for computing the holographic entanglement entropy of a boundary strip aligned along the $x^i$-direction (with the corresponding extremal surface illustrated in Fig.\,\ref{FigureEntanglementEntropy}) is as follows:
        \begin{enumerate}
            \item Select a turning point $r_*$ sufficiently close to the horizon to ensure the extremal surface probes the IR geometry.
            \item Compute the conserved quantity $C_i$ for the extremal surface using  \eqref{ExtremalSurfaceConstant}.
            \item Determine the corresponding boundary strip width $l_i$ by evaluating the integral in \eqref{StripWidth}.
            \item Adjust the value of $r_*$ iteratively until the obtained width $l_i$ matches the desired scale for physical analysis.
            \item Finally, compute the holographic entanglement entropy via \eqref{StripEntanglementEntropy}.
        \end{enumerate}
        
        Following the discussions in \cite{DifferentialEntropyBalasubramanian_2014,DifferentialEntropyMyers_2012,DifferentialEntropyLiu_2014,CFunctionOrderParameterBaggioli_2023}, the conserved constant $C_i$ is identified with the rate of {the first derivative of the entanglement entropy}. Specifically
        \begin{equation}
            \label{StripDifferentialEntropy}
            \frac{4G}{L^2}\frac{\partial S_i}{\partial l_i}=\frac{C_i}{2},
        \end{equation}
        and this procedure yields the c-function, which encodes the renormalization group flow in the dual field theory
        \begin{equation}
            \label{HolographicCFunction}
            c_i=\frac{4G}{L^2}\frac{\partial S_i}{\partial l_i}{l_i}^3=\frac{C_i}{2}{l_i}^3.
        \end{equation}

        \begin{figure}[htbp]
            \begin{minipage}{0.49\linewidth}
                \centering
                \includegraphics[width=\textwidth]{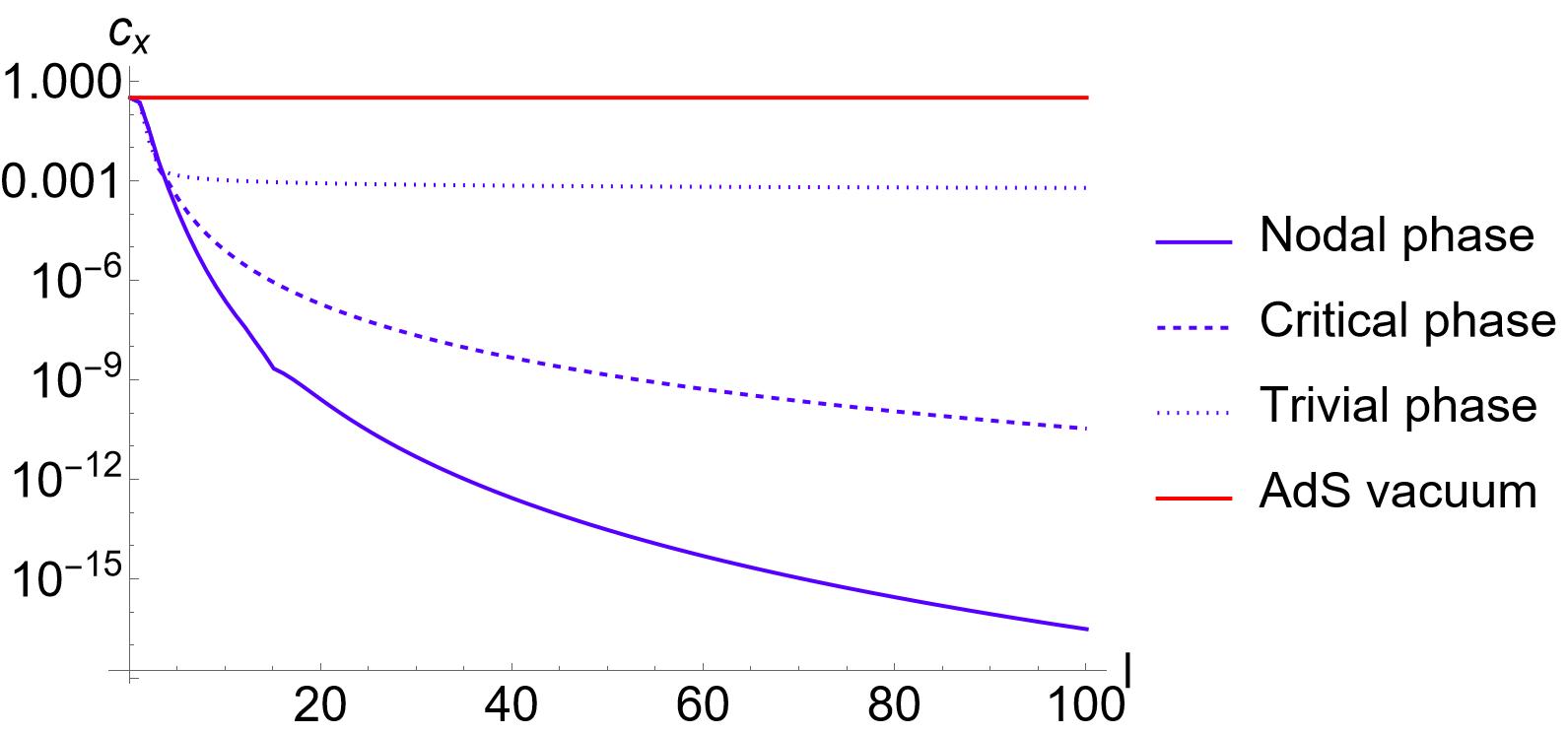}
            \end{minipage}
            \begin{minipage}{0.49\linewidth}
                \centering
                \includegraphics[width=\textwidth]{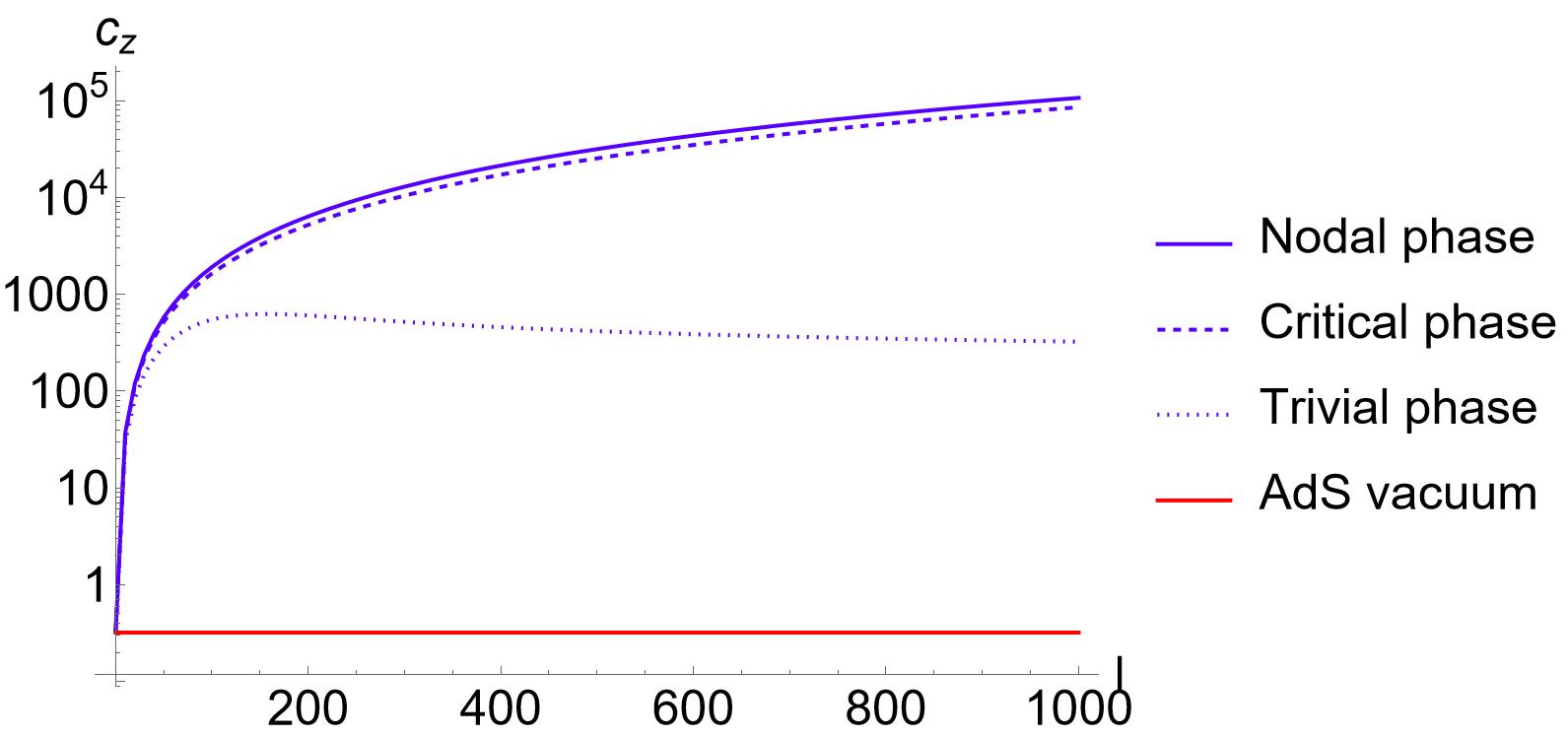}
            \end{minipage}
            \caption{Left: The evolution of the holographic c-function $c_x$ with $l_x$ for different phases. Right: The evolution of the holographic c-function $c_z$ with $l_z$ for different phases.}
            \label{Figurel-cfunction}
        \end{figure}

        We have three c-functions, $c_x,~c_y$ and $c_z$. Because of the rotational symmetry on the $xy$-plane, $c_x$ is equal to $c_y$, so we only consider $c_x$ and $c_z$ without loss of generality.
The c-theorem dictates that in isotropic, Lorentz-invariant theories, the c-function must decrease monotonically with the spatial scale $l$, reflecting the irreversible coarse-graining of degrees of freedom under the RG flow from the UV to the IR. In anisotropic systems, however, this monotonicity is no longer guaranteed\cite{Chu:2019uoh}. As shown in Fig.\,\ref{Figurel-cfunction}, the behavior of the holographic c-function for a nodal-line semimetal depends sensitively on direction. We plot $c_i(l_i)$ as a function of the strip width $l_i$
  separately for the $x$- and $z$- directions, across the topologically nontrivial, trivial, and critical phases, as well as for the pure AdS background. 
  
  In Fig.\,\ref{Figurel-cfunction}, we note that as $l$ increases, the c-function in the x-direction ($c_x$) decreases for all non-vacuum phases, whereas in the z-direction ($c_z$) it increases.
  When the strip is aligned along the $x$-direction, due to the behavior of $f(r)$, the RT surface becomes pinched as it probes deeper into the IR geometry, resulting in a growth rate of its area that is lower than that in pure AdS. Consequently, $c_x$ decreases with increasing $l_x$. Conversely, when the strip is aligned along the $z$-direction, the contraction governed by $u(r)$ is less severe, leading to an area growth rate that exceeds the pure AdS case, and thus $c_z$ increases with $l_z$. This reflects the underlying anisotropy of the nodal line semimetal.

At small $l$, the behavior of the c-function is governed by the UV geometry of the bulk spacetime, which exhibits the same asymptotic AdS structure for all backgrounds considered, namely, the topologically nontrivial, trivial, and critical phases, as well as the pure AdS reference case. In contrast, its large $l$ behavior is dictated by the distinct IR geometry associated with each phase. It can be seen that in pure AdS spacetime, the c-function remains constant with increasing $l$, whereas the topologically trivial phase stays close to the pure AdS case, and the topologically non-trivial phase deviates more significantly. It is the different IR scaling behaviors that lead to the markedly different scaling regimes of the c-function at large strip widths, as detailed below.

        For the c-function, a power-law dependence emerges at large $l$. Specifically, for the topologically nontrivial phase, the critical phase, the topologically trivial phase, and the AdS vacuum, the exponent $\mathbf{z}$ takes the values 10.929, 6.3694, 1, and 1, respectively. When the strip is aligned along the x-direction, $c_x$ scales as $l_x^{1-\mathbf{z}}$; when it is along the z-direction, $c_z$ scales as $l_z^{2-\frac{2}{\mathbf{z}}}$. This explains the distinct behaviors of the c-function among the phases observed in Fig.\,\ref{Figurel-cfunction}. Moreover, because the scaling behavior differs from one phase to another, a sharp transition at the critical point naturally appears.

        \begin{figure}[htbp]
            \begin{minipage}{0.49\linewidth}
                \centering
                \includegraphics[width=\textwidth]{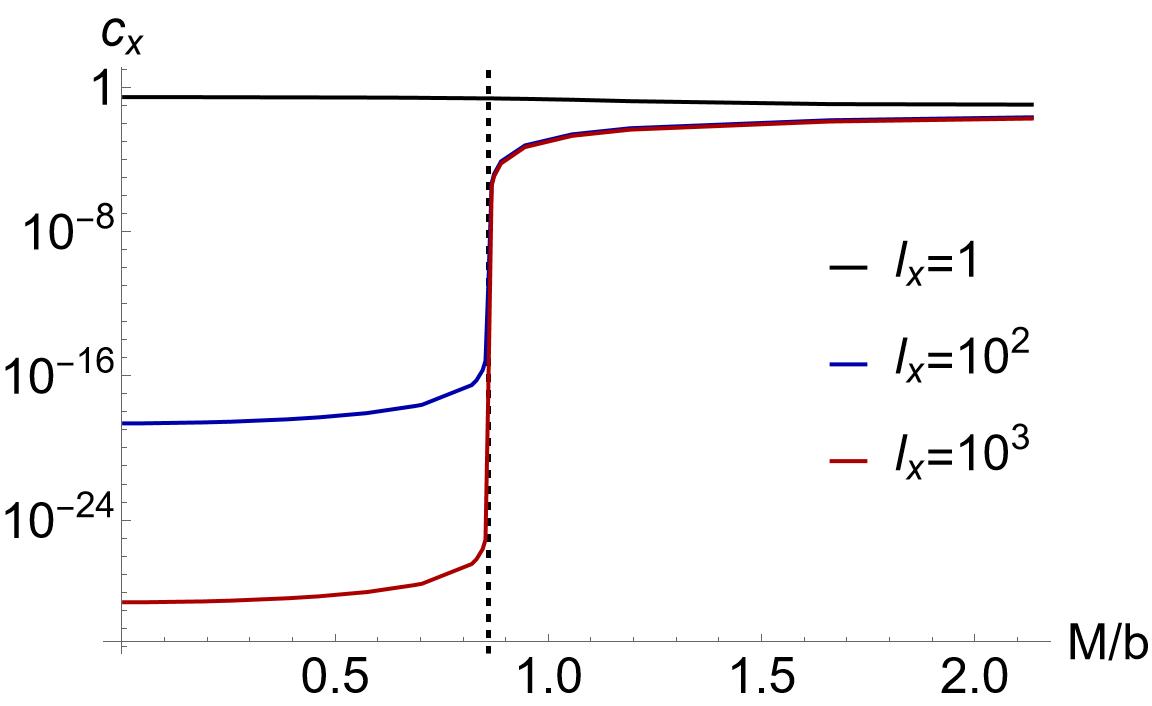}
            \end{minipage}
            \begin{minipage}{0.49\linewidth}
                \centering
                \includegraphics[width=\textwidth]{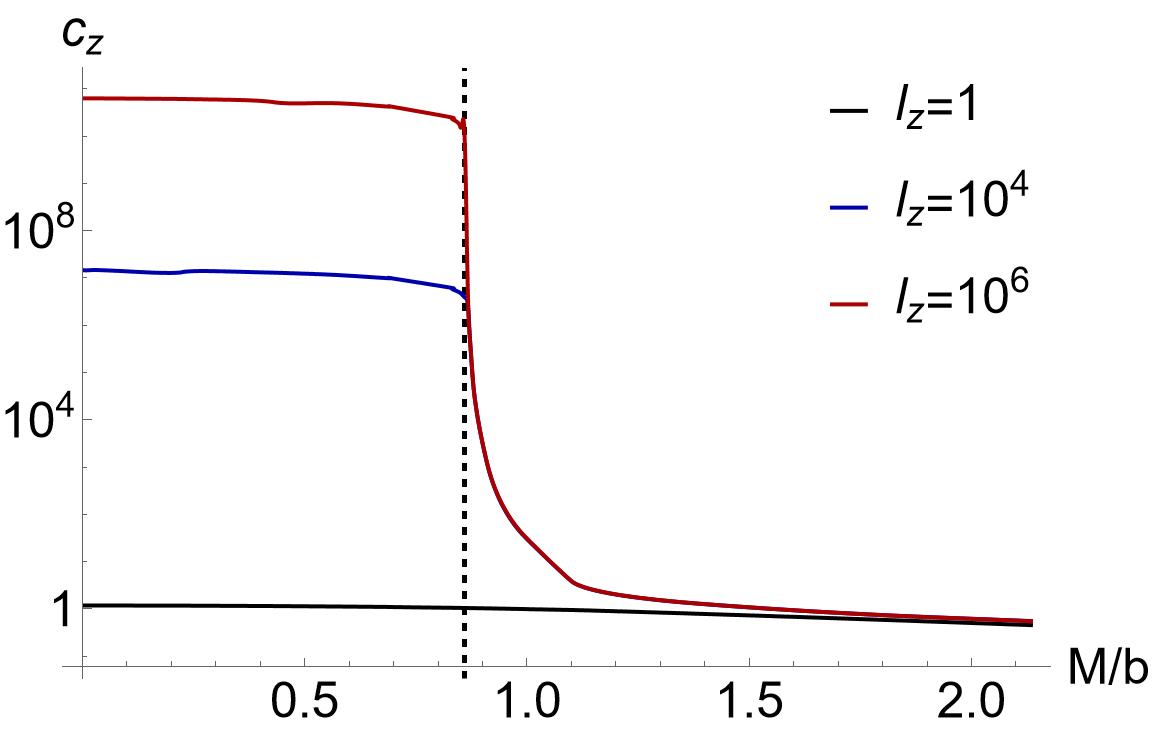}
            \end{minipage}
            \caption{Left: the evolution for $c_x$ with increasing $M/b$. Right: the evolution for $c_z$ with increasing $M/b$.}
            \label{FigureCFunction}
        \end{figure}

            Therefore, the distinct large $l$ scaling of the c-function serves as a direct probe of the low-energy IR physics characteristic of each phase. This makes the large $l$ value of the c-function a powerful non-local order parameter for detecting topological quantum phase transitions. Fig.\,\ref{FigureCFunction} depicts the dependence of the c-functions at large values of $l$ on values of $M/b$. In Fig.\,\ref{FigureCFunction}, we note that when the strip width is chosen to be very small, the c-function is approximately constant and does not vary with $M/b$. This aligns with our expectation: for small strip widths, the corresponding RT surface cannot probe deep into the IR region and only captures the vacuum contribution of the asymptotic AdS spacetime. This indicates that to clearly reveal changes in the entanglement structure across the topological phase transition, we must always choose a sufficiently large boundary strip width to ensure that the corresponding RT surface can probe the IR geometry.
            
            As shown in Fig.\,\ref{FigureCFunction}, the c-function evaluated at sufficiently large $l$ depends sharply on the dimensionless parameter $M/b$, exhibiting a clear discontinuity or singular feature precisely at the critical point. This behavior  confirms that the holographic c-function can distinguish between topologically trivial and nontrivial phases, providing a sharp diagnostic of the transition. In the next sections, we will study three different types of tripartite entanglement measures for the holographic nodal line semimetal and utilize their large $l$ behaviors as a probe for the quantum topological phase transitions. 

\section{The conditional mutual information and its  scaling behavior}
\label{Section3}
    Having established the holographic c-function as a sensitive probe of the topological quantum phase transition in nodal-line semimetals, we now turn to more complicated multipartite entanglement structures. While bipartite measures like entanglement entropy capture essential data, the rich physics of strongly coupled topological phases is often encoded in more complex, multi-party entanglement patterns. Investigating these patterns is crucial for a complete understanding of properties of the ground state and topological phase transitions in strongly coupled holographic semimetal systems.

    Motivated by these considerations, we study the conditional mutual information, a tripartite entanglement measure that has been widely used in condensed matter systems to characterize correlation structures beyond bipartite entanglement and has been studied extensively in holography \cite{DifferentialEntropyBalasubramanian_2014,DifferentialEntropyLiu_2014,DifferentialEntropyMyers_2012}.  CMI quantifies the correlations between two subsystems conditioned on a third. For our holographic setup, we will compute the CMI for a specific boundary configuration of two infinitesimal strips separarted at a distance $l$ with the condition being the strip region in between. Analogous to the c-function, we will also examine its large $l$ scaling behavior. Since this IR physics is governed by the low-energy fixed point geometry, the distinctive large $l$ scaling behaviors of CMI in different phases can serve as a robust, non-local diagnostic of the topological quantum phase transition. We will demonstrate that the CMI indeed exhibits a sharp feature at the critical point, confirming its role as another powerful entanglement-based order parameter for the transition.

    \subsection{Setup and calculations for CMI}       
        The conditional mutual information $I(A:B|E)$ quantifies the bipartite entanglement between $A$ and $B$ conditioned on $E$. It is defined as
        \begin{equation}
            \label{ConditionalMutualInformation}
            I(A:B|E)=S(A\cup E)+S(B\cup E)-S(A\cup B\cup E)-S(E),
        \end{equation}
        which measures the information between $A$ and $B$ when $E$ is known.

        In condensed matter physics, the conditional mutual information has emerged as a key diagnostic for distinguishing between short-range entangled and long-range entangled states, the latter being the hallmark of intrinsic topological order \cite{Wen:2012hm}. The CMI has contributions from tripartite entanglement among the three subsystems and it is sensitive to the non-local entanglement structure that underpins topological phases \cite{Kato:2016lre, QuantumInformationCondensedMatterZeng_2019}. Simultaneously, the CMI serves as a crucial tool for analyzing the entanglement structures in holographic systems \cite{ MultiEntanglementUpperBoundju2024holographicmultipartiteentanglementupper, ConditionalMutualInformationJu:2024kuc}, and it provides a direct link between boundary entanglement and the geometry of the bulk spacetime \cite{DifferentialEntropyMyers_2012, ju2024generalizedrindlerwedgeholographic,DifferentialEntropyBalasubramanian_2014, ConditionalMutualInformationJu_2024, ConditionalMutualInformationJi_2025}.
        
        Motivated by these developments, we study the CMI in holographic nodal line semimetals as a probe of their  multipartite entanglement structures. We consider a geometric configuration consisting of two infinitesimal boundary regions separated by a finite-width strip, with the strip taken as the conditioning region. In this setup, the separation scale $l$ controls the depth to which the associated RT surfaces probe the bulk geometry, so that the large $l$ behavior of the CMI is directly governed by the infrared region of the dual spacetime.  Physically, this configuration isolates multipartite correlations mediated through the intermediate region and is therefore well suited to diagnosing long-range multipartite entanglement for the corresponding states. Technically, the use of infinitesimal regions together with translational symmetry allows the CMI to be expressed in terms of strip entanglement entropies, rendering the holographic computation tractable.

        As shown in \cite{ConditionalMutualInformationJu_2024,EntanglementContourChen_2014}, the conditional mutual information for two infinitesimal regions $A$ and $B$, separated by a strip $E$ of width $l$, can be expressed as the second derivative of the entanglement entropy:
        \begin{equation}
            I(A:B|E)=-\frac{d^2S}{dl^2},
        \end{equation}
        where $S$ is the holographic entanglement for strip $E$. Combining this with the relation between the entropy derivative and the conserved quantity given in \eqref{StripDifferentialEntropy}, the CMI can be recast directly in terms of the conserved constant $C_i$ for the corresponding RT surface
        \begin{equation}
            \label{HolographicCMI}
            I(A:B|E)=-\frac{L^2}{8G}\frac{dC_i}{dl_i},
        \end{equation}
        where $C_i$ is defined in \eqref{ExtremalSurfaceConstant} and $l_i$ is the width for the strip aligned along the $x^i$-direction. The coefficient $L^2/(8G)$, involving the transverse cutoff $L$ and the gravitational constant $G$, is often omitted when comparing the functional behavior of the CMI across different phases, as it amounts to an overall constant scaling. Therefore, the essential quantity we compute is proportional to $-\frac{dC_i}{dl_i}$. In summary, the protocol for calculating the CMI for two infinitesimal subregions seperated by a strip aligned along the $x^i$-direction is the following
        \begin{enumerate}
            \item Select the turning point $r_*$ sufficiently close to the horizon and compute the conserved constant $C_i=\sqrt{g_{xx}(r_*)g_{yy}(r_*)g_{zz}(r_*)}$.
            \item Determine the associated boundary strip width $l_i$ by evaluating the integral in \eqref{StripWidth}.
            \item Perform a numerical derivative of $C_i$  with respect to $l_i$  to obtain the core quantity $-dC_i/dl_i$, which is proportional to the CMI via \eqref{HolographicCMI}.
            \item Adjust the value for $r_*$ to make sure the width $l_i$ matches the desired scale.
        \end{enumerate}
    \subsection{The large \texorpdfstring{$l$}{\textit{l}} behavior as a probe of quantum topological phase transitions}    
        \begin{figure}[htbp]
            \begin{minipage}{0.49\linewidth}
                \centering
                \includegraphics[width=\textwidth]{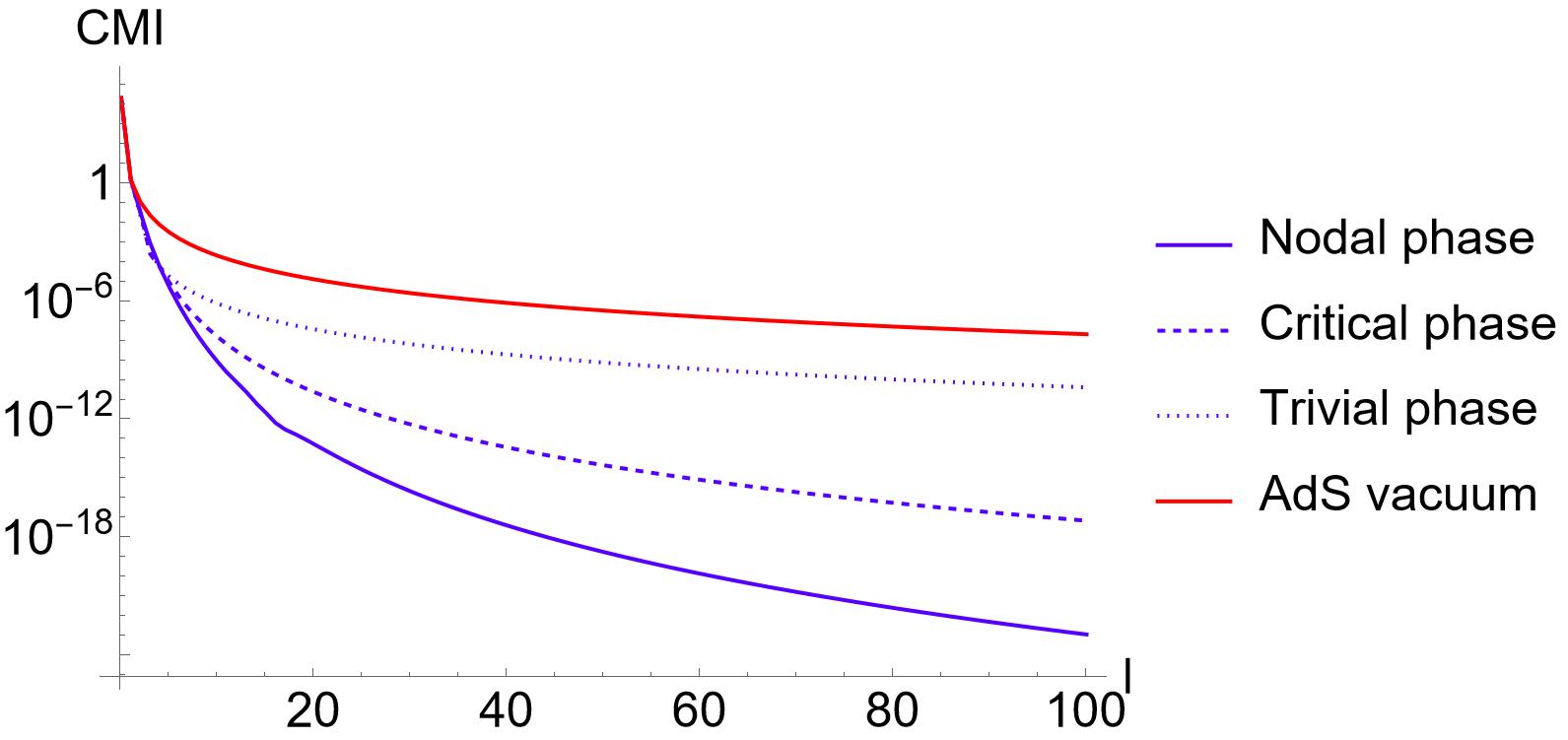}
            \end{minipage}
            \begin{minipage}{0.49\linewidth}
                \centering
                \includegraphics[width=\textwidth]{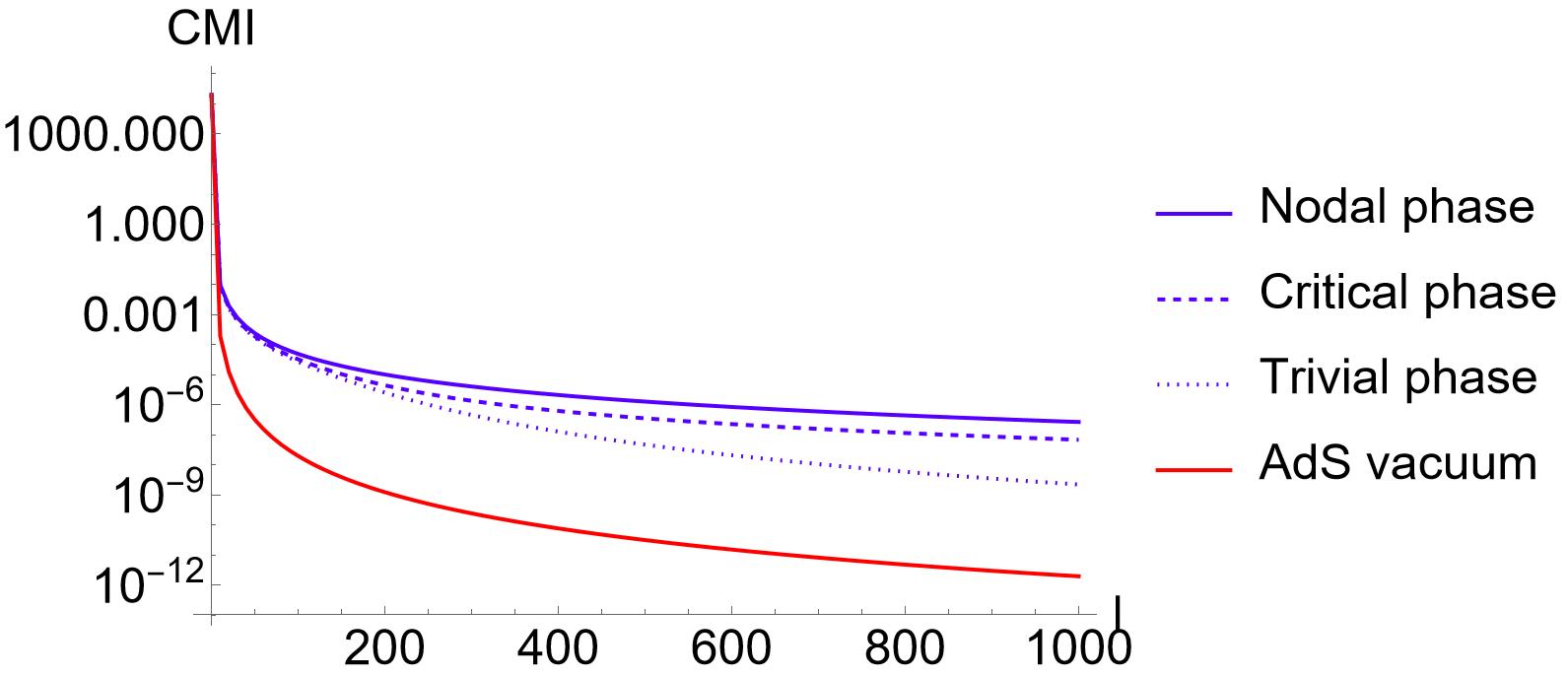}
            \end{minipage}
            \caption{Left: The dependence of the values of CMI $I(A:B|E)$ on $l_x$ (left) and $l_z$ (right) for different phases with representative values of $M/b$ in each phase.}
            \label{Figurel-CMI}
        \end{figure}
        
        Fig.\,\ref{Figurel-CMI} displays the value of the CMI as a function of the strip width $l$ for different phases as well as for the pure AdS background. The value of CMI at $l\to \infty$ vanishes for all phases and both in the $x$- and the $z$-directions. This confirms that the strongly coupled holographic nodal line semimetal is still a short range entangled state (SRE), while not a long range entangled state (LRE). One can see that in the $x$-direction the CMI of the topological non-trivial phase lies below that of the pure AdS state; as the system undergoes the phase transition toward the topologically trivial phase, the curve gradually approaches the pure AdS curve. In the $z$-direction the opposite occurs: the CMI of the topologically non-trivial phase exceeds the value for pure AdS, and as the system becomes trivial, it gradually decreases toward the AdS vacuum curve.

        This behavior again reflects the anisotropy of the system. This shows that the nodal ring in the $k_x-k_y$ plane suppresses long range correlations along the $x$ and $y$-directions, while enhancing long range correlation along the $z$-direction. This observation is consistent with the behavior previously deduced from the c-function calculation, namely that along the $x,y$-directions degrees of freedom for the system freeze out along the renormalization group flow, driving the system toward the topologically nontrivial phase, whereas along the $z$-direction a larger number of degrees of freedom remain active.

\begin{figure}[htbp]
            \begin{minipage}{0.49\linewidth}
                \centering
                \includegraphics[width=\textwidth]{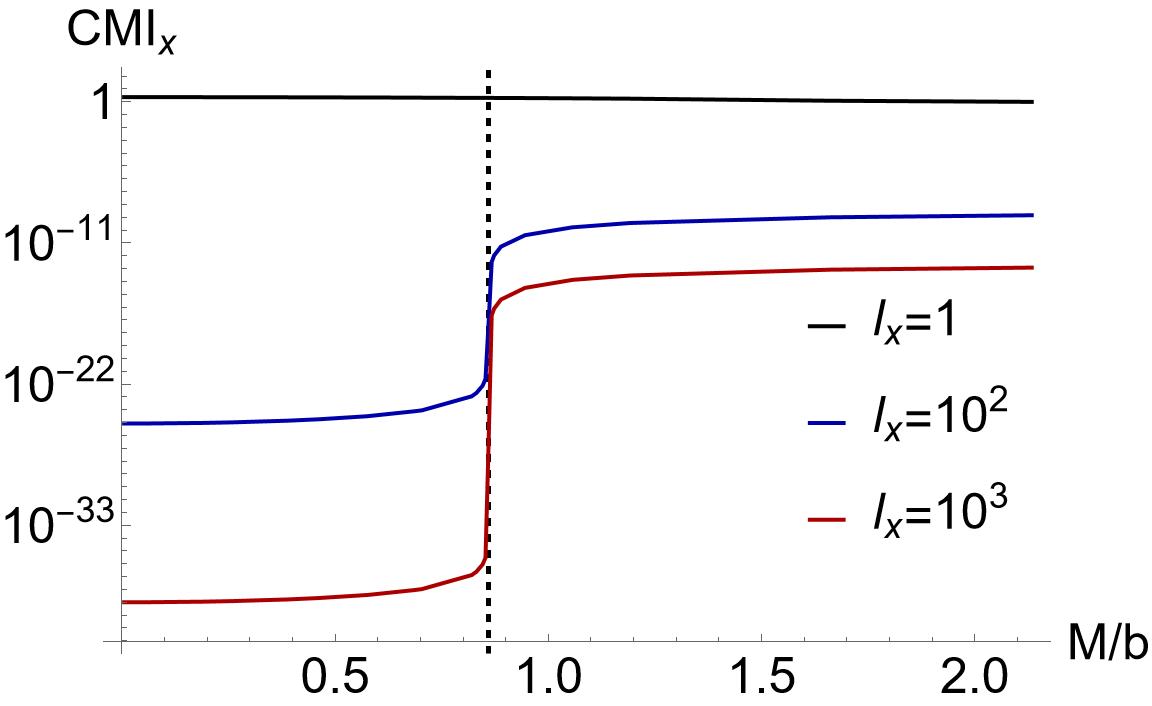}
            \end{minipage}
            \begin{minipage}{0.49\linewidth}
                \centering
                \includegraphics[width=\textwidth]{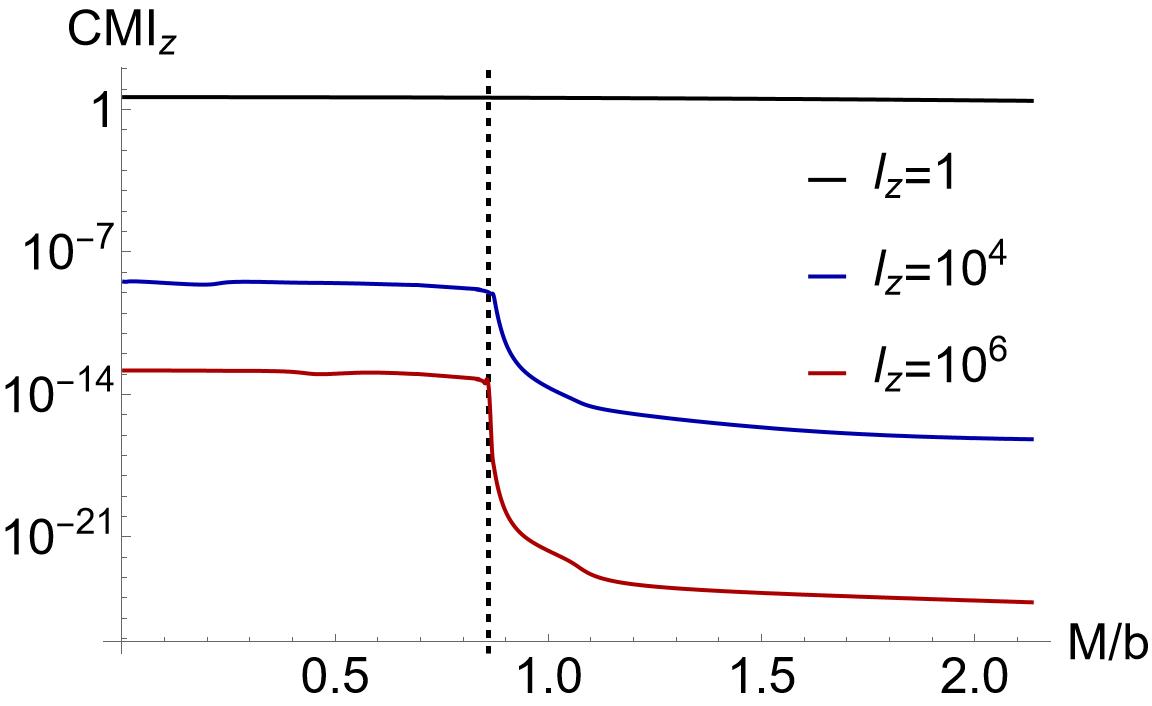}
            \end{minipage}
            \caption{Left: the evolution for the CMI with increasing $M/b$, where the strip is aligned along the $x$ direction. Right: the evolution for the CMI with increasing $M/b$, where the strip is aligned along the $z$ direction.}
            \label{FigureCMI}
        \end{figure}

        The scaling behavior of CMI at large $l$ reflects the IR scaling of various phases and could be utilized as a probe for the corresponding quantu m phase transition. We find the following scaling exponents for the large $l$ dependence of CMI. When the strip is aligned along the $x$-direction, corresponding CMI scales as $l_x^{-3-\mathbf{z}}$; when the strip is along the $z$-direction, corresponding CMI scales as $l_z^{-2-\frac{2}{\mathbf{z}}}$, where $\mathbf{z}$ takes the values 10.929, 6.3694, 1, and 1 as defined in \eqref{InfraRedScale} for the topologically nontrivial phase, the critical phase, the topologically trivial phase, and the AdS vacuum respectively. This explains the distinct behaviors of the CMI among the phases observed in Fig.\,\ref{Figurel-CMI}. Moreover, because the scaling behavior differs from one phase to another, a sharp transition at the critical point naturally appears as shown in Fig.\,\ref{FigureCMI}.   

        As the large $l$ behavior is determined by the IR scaling properties of each phase, we could use the large $l$ values of CMI as a probe of quantum phase transitions. Fig.\,\ref{FigureCMI} shows the values of CMI as a function of $M/b$ at several fixed values of $l_x$ and $l_z$. We can see that at small strip widths, the conditional mutual information of different phases exhibits almost no variation. This is because, for small strip widths, the corresponding RT surfaces cannot probe deep into the IR region and only access the UV region of the asymptotic AdS spacetime, which is similar for all phases. When the strip width becomes large, the RT surfaces penetrate deeper into the bulk geometry, enabling us to observe a jump in the CMI at the critical point. As the strip width increases, the CMI of every phase decreases, reflecting the weakening correlation between the two subsystems separated by the strip as they become farther apart. It is worth to also note that in the $x$-direction the CMI increases with $M/b$, whereas in the $z$-direction it decreases with $M/b$. 
        
        As shown in Fig.\,\ref{FigureCMI}, with increasing $M/b$, the CMI corresponding to each direction truly exhibits a sharp transition at the critical point for large enough $l$. This confirms that the values of CMI at large $l$ can indeed be used to characterize the quantum phase transition between topological trivial and non-trivial phases as a new non-local order parameter.

\section{Multipartite entanglement from the holographic multi-entropy and the scaling behavior}
    \label{Section4}
    In the previous sections, we analyzed the entanglement structure of holographic nodal line semimetals using entanglement measures constructed from entanglement entropies, namely the c-function and the conditional mutual information. In this section, we turn to a distinct class of multipartite entanglement measures based on a new quantity:  the multi-entropy. The multi-entropy has been proposed as a natural extension of entanglement entropy to multiple disjoint regions and provides a refined characterization of multipartite entanglement patterns \cite{MultiEntropyHarper_2024,Balasubramanian2026,jiang2025holographicdualghzstate,Yuan2025}. In holographic systems, it admits a geometric realization in terms of extremal bulk networks, such as Steiner trees, anchored to the chosen boundary regions. We will consider a special genuine multipartite measure, $\kappa$, constructed from the multi-entropy, which is expected to capture genuinely tripartite entanglement structures that are not reducible to bipartite entanglement.

In this section, we first briefly review the definition of multi-entropy and its holographic formulation. We then compute the holographic multi-entropy and the tripartite measure $\kappa$ for nodal line semimetals and analyze the scaling behavior at large separation length
$l$. We show that the large $l$ behavior of $\kappa$ is governed by the IR geometry and undergoes a sharp change at the quantum topological phase transition, demonstrating that it provides an effective multipartite entanglement probe of the phase structure.
    
    \subsection{The holographic multi-entropy}
        The holographic multi-entropy generalizes the Rényi entropy to multipartite scenarios via twist operators with higher-genus monodromy structures.  In what follows, we first review the definition of Rényi entropy via the replica trick, then extend this formalism to define the multi-entropy, and finally compute it within our holographic nodal-line semimetal model \eqref{NodalHolographicModel}.
        
        The entanglement entropy for a boundary subregion $A$, defined as $S_A=-{\rm Tr\,}\rho\log\rho$, is often difficult to compute directly. This motivates the use of Rényi entropy $S_n=\frac{1}{1-n}\log({\rm Tr\,} \rho^n)$, which reduce to $S_A$ in the limit $n\to 1$. More importantly, the Rényi construction admits a natural generalization to multipartite entanglement via twist operators associated with higher monodromy groups. Specifically, for a $q$-partite decomposition, one defines the multi-Rényi entropy for the $q$-partite pure state $\ket{\psi}$ as
        \begin{equation}
            \label{MultiRenyiEntropy}
            \begin{aligned}
                S_n^{(q)}&=\frac{1}{1-n}\frac{1}{n^{q-2}}\log\frac{Z_n^{(q)}}{\left(Z_1^{(q)}\right)^{n^{q-1}}},\\
                Z_n^{(q)}&=\bra{\psi}^{\otimes n^{q-1}}\sigma_{1}(g_1)\sigma_{2}(g_2)\cdots\sigma_{q}(g_q)\ket{\psi}^{\otimes n^{q-1}},
            \end{aligned}
        \end{equation}
        where each twist operator $\sigma_{i}(g_i)$ corresponds to an element of $\mathbb{Z}_n^q$ with the constraint $\prod_ig_i=1$. The multi-entropy $S^{(q)}$ is then obtained by taking the limit $\lim_{n\to 1}S_n^{(q)}$. This framework extends the bipartite entanglement measures to multipartite correlations, which will be employed to probe the entanglement structure of the holographic nodal line semimetal.
        
        Within the framework of holographic duality, consider the boundary partitioned into $q$ connected regions, denoted collectively as $A_i,~i=1,q$.  The holographic dual of the multi-entropy is given by a Steiner tree in the bulk \cite{SteinerTreeAlkalaev2019,SteinerTreeLunin2001,MultiEntropyHarper_2024,Gadde_2022,Gadde_2023,gadde2025multiinvariantsbulkreplicasymmetry}. This tree is composed of a network of minimal surfaces $\{\Gamma_i\}$ which collectively partition the bulk into $q$ regions, each of which is homologous to a distinct boundary subregion $A_i$. Crucially, this network may include internal junction points, lines, or higher-dimensional surfaces depending on the bulk dimensionality—where several minimal surfaces meet. Among all such admissible networks, the relevant Steiner tree is the minimal-area one. The holographic multi-entropy is then geometrized by the total area of this minimal Steiner tree
        \begin{equation}
            S^{(q)}=\frac{1}{4G}({\text{minimal total area for the Steiner tree}}).
        \end{equation}
        
\begin{figure}[htbp]
            \centering
            \includegraphics[width=0.6\textwidth]{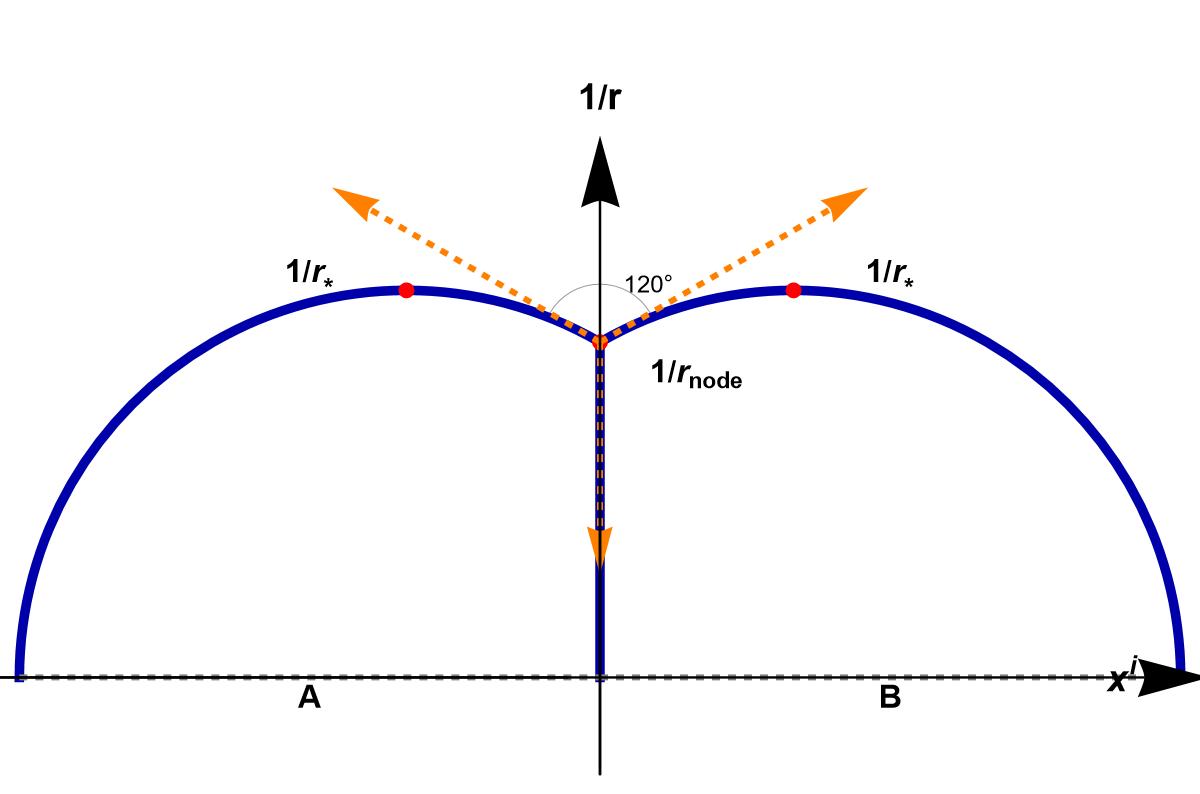}
            \caption{The Steiner tree on a slice $x^j(j\ne i)$ for two strips $A,~B$ aligned along $x^i$ direction with the same width and their complement $(A\cup B)^c$. Three minimal surfaces have the junction points denoted as $r_{\text{node}}$, and they meet at mutual angel of $2\pi/3$.}
            \label{MultiEntropySteinTree}
        \end{figure}

        In the following we will focus on tripartite configurations with $q=3$. To compute the holographic $3$-partite multi-entropy for two parallel strips, $A$ and $B$, aligned along the $x^i$-direction and their complement $(A\cup B)^c$ in the nodal line semimetal model \eqref{NodalHolographicModel}, we assume, for simplicity, that $A$ and $B$ have equal width. Owing to translational symmetry in the transverse directions $x^j(j\ne i)$, the Steiner tree is invariant on slices of constant $x^j(j\ne i)$. As shown in Fig.\,\ref{MultiEntropySteinTree}, on each such slice it consists of three curves that meet at a common junction point located at a radial coordinate $r_{\text{node}}$. Each curve is a minimal geodesic derived from the reduced action \eqref{ExtremalSurfaceAction}. The minimality of the total area imposes a geometric constraint at the junction: the three curves must meet at mutual angles of $2\pi/3$ \sun{\cite{MultiEntropyHarper_2024}}. This is the familiar equilibrium condition for a Steiner network minimizing the total length in a Riemannian geometry.
        
        The conserved constant $C_i$ , defined in \eqref{ExtremalSurfaceConstant}, is a constant along each minimal geodesic. Therefore, its value at the junction point (radial coordinate $r_{\text{node}}$) must equal its value at the corresponding turning point (radial coordinate $r_*$) on each leg:
        \begin{equation}
            \sqrt{g_{xx}(r_*)g_{yy}(r_*)g_{zz}(r_*)}=\sqrt{\frac{g_{ii}(r_{\text{node}})}{g_{ii}(r_{\text{node}})+g_{rr}(r_{\text{node}})r'^2}}\sqrt{g_{xx}(r_{\text{node}})g_{yy}(r_{\text{node}})g_{zz}(r_{\text{node}})}.
        \end{equation}
        
        The equilibrium condition at the junction, arising from the minimal area requirement, imposes the geometric constraint $3g_{rr}(r_{\text{node}})r'^2=g_{ii}(r_{\text{node}})$. Substituting this into the equation above yields a direct relation between $r_*$ and $r_{\text{node}}$: $\sqrt{g_{xx}(r_*)g_{yy}(r_*)g_{zz}(r_*)}=\frac{\sqrt{3}}{2}\sqrt{g_{xx}(r_{\text{node}})g_{yy}(r_{\text{node}})g_{zz}(r_{\text{node}})}$. This relation allows us to determine the turning point $r_*$ from a chosen junction depth $r_{\text{node}}$. Subsequently, the holographic multi-entropy for two adjacent strips of equal width and their complement can be computed
        \begin{equation}
            \label{HolographicMultiEntropy}
            \begin{aligned}
                \text{Multi-Entropy}&=2\int_{r_*}^\infty\prod_{j=1}^n g_{jj}\sqrt{\frac{g_{rr}}{g_{ii}(\prod_{j=1}^n g_{jj}-{C_i}^2)}}dr\\
                &+2\int_{r_*}^{r_{\text{node}}}\prod_{j=1}^n g_{jj}\sqrt{\frac{g_{rr}}{g_{ii}(\prod_{j=1}^n g_{jj}-{C_i}^2)}}dr+\int_{r_{\text{node}}}^{\infty}\sqrt{g_{rr}\prod_{j\ne i}g_{jj}}dr.
            \end{aligned}
        \end{equation}
        
        In summary, the protocol for computing the multi-entropy for two strip $A,~B$ aligned along the $x^i$-direction with the same width and their complement $(A\cup B)^c$ in the holographic nodal line semimetal is following:
        \begin{enumerate}
            \item Select the junction point $r_{\text{node}}$ sufficiently close to the horizon, compute the turning point $r_*$ via solving the equation $$\sqrt{g_{xx}(r_*)g_{yy}(r_*)g_{zz}(r_*)}=\frac{\sqrt{3}}{2}\sqrt{g_{xx}(r_{\text{node}})g_{yy}(r_{\text{node}})g_{zz}(r_{\text{node}})},$$ 
            and compute the conserved constant $C_i=\sqrt{g_{xx}(r_*)g_{yy}(r_*)g_{zz}(r_*)}$.
            \item Determine the associated boundary strip width $l_i$ for the subregion $A$ or $B$ by evaluating the integral
            $$l_i=\int_{r_{*}}^\infty \sqrt{\frac{g_{rr}{C_i}^2}{g_{ii}(\prod_{j=1}^n g_{jj}-{C_i}^2)}}dr+\int_{r_{*}}^{r_{\text{node}}} \sqrt{\frac{g_{rr}{C_i}^2}{g_{ii}(\prod_{j=1}^n g_{jj}-{C_i}^2)}}dr.$$
            \item Adjust the value for $r_{\text{node}}$ to make sure the width $l_i$ matches the desired scale.
            \item Computing the multi-entropy for two strip $A,~B$ and their complement $(A\cup B)^c$ by evaluating the integral in \eqref{HolographicMultiEntropy}.
        \end{enumerate}
        
        Using this procedure, we can compute the multi entropy for two strips $A,~B$ and their complement. The multi-entropy is a UV-divergent quantity whose dominant contribution comes from the asymptotic AdS UV region, which overwhelms any IR contribution. Therefore, the resulting value is insensitive to the strip orientation, strip width, and the phase of the system.
        
        We need to employ a subtraction procedure to remove the UV contribution and obtain a finite quantity and this would also help isolate genuine tripartite entanglement from bipartite entanglement.
    \subsection{\texorpdfstring{$\kappa$}{\textit{kappa}}: a multipartite entanglement measure from multi-entropy}
    A generic feature of multipartite entanglement measures, such as the multi-entropy, is that they capture the total entanglement among all subsystems, including both bipartite and genuine multipartite contributions. Taking the tripartite case as an example, for a boundary divided into regions $A,~B$, and $C$, the multi-entropy $S^{(3)}(A:B:C)$ includes not only the irreducible tripartite entanglement but also the bipartite contributions between each pair ($A\leftrightarrow B$, $B\leftrightarrow C$ and $C\leftrightarrow A$). To isolate the genuinely tripartite component, we subtract the summed bipartite entropies. This subtraction is equivalent to extracting the finite, universal part from the UV-divergent multi-entropy, which yields the definition
        \begin{equation}
            \label{Kappa}
            \kappa(A:B:C)=S^{(3)}(A:B:C)-\frac{1}{2}(S_{AB}+S_{BC}+S_{CA}).
        \end{equation}
        
        Thus defined, $\kappa$ is an infrared quantity insensitive to ultraviolet cutoffs. It subtracts off all bipartite contributions and isolates the genuine tripartite entanglement that cannot be decomposed into bipartite correlations. As further demonstrated in \cite{GenuineMulipartiteiizuka2025genuinemultientropyholography,GenuineMulipartiteiizuka2025genuinemultientropy}, $\kappa$ vanishes for separable states and triangle states, the latter defined as
        \begin{equation}
            \label{DefinitionTriangleState}
            \ket{\psi}_{ABC}=\ket{\psi}_{A_LB_R}\ket{\psi}_{B_LC_R}\ket{\psi}_{C_LA_R},
        \end{equation}
        for some appropriate bipartition $\mathcal{H}_{\alpha}=\mathcal{H}_{\alpha_L}\otimes\mathcal{H}_{\alpha_R},\alpha=A,~B$ and $C$. We can see in this expression that a triangle state has tripartite entanglement that could be reduced to bipartite entanglement among smaller subsystems of the original three subsystems. Therefore, in a triangle state, there is no genuine tripartite entanglement. As $\kappa$ vanishes for a triangle state, while $\kappa$ is non-zero for non-triangle states such as the GHZ state, $\kappa$ is therefore expected to be a genuine multipartite entanglement measure.
    \subsection{Large \texorpdfstring{$l$}{\textit{l}} behavior of \texorpdfstring{$\kappa$}{\textit{kappa}} as a probe of quantum topological phase transitions}
\begin{figure}[htbp]
            \begin{minipage}{0.49\linewidth}
                \centering
                \includegraphics[width=\textwidth]{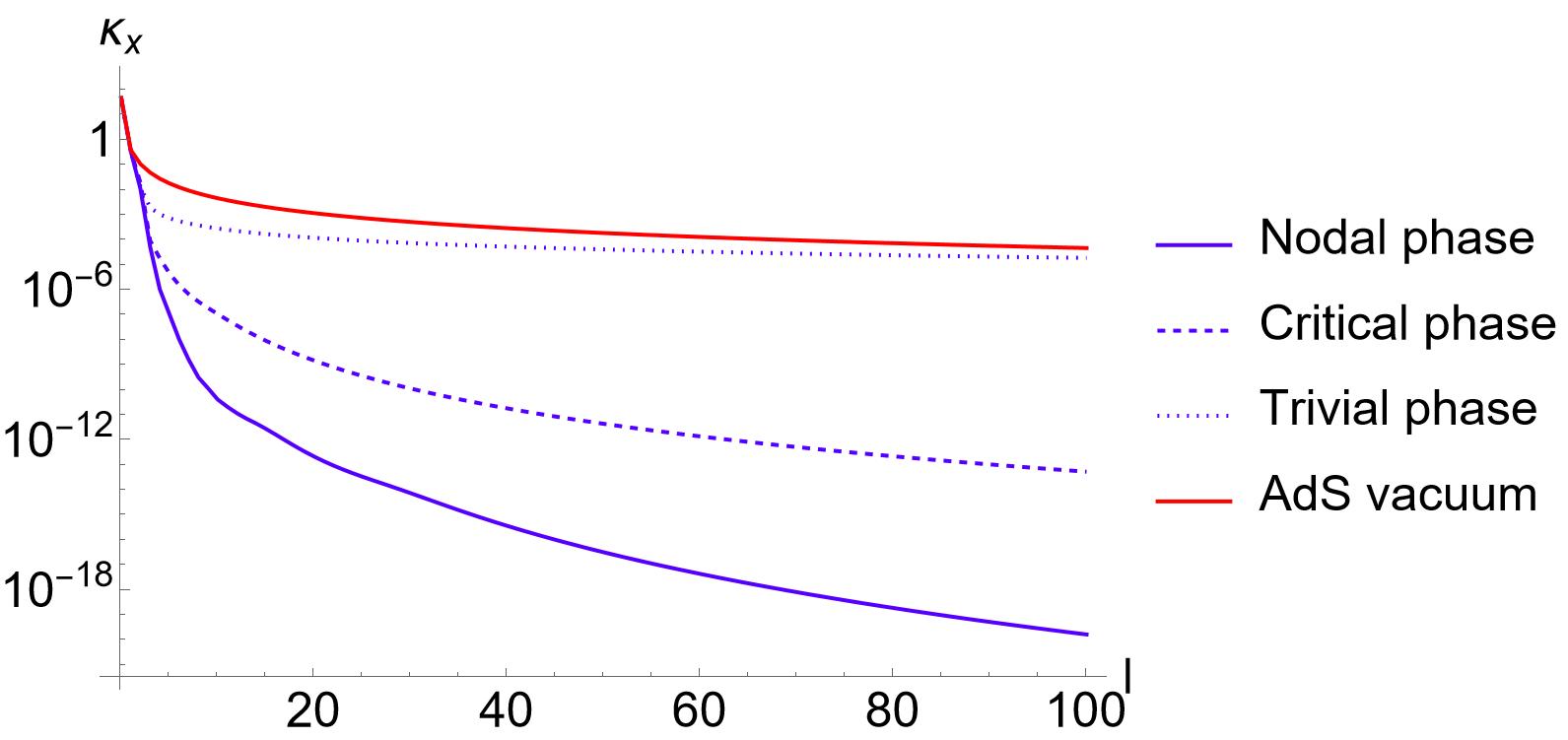}
            \end{minipage}
            \begin{minipage}{0.49\linewidth}
                \centering
                \includegraphics[width=\textwidth]{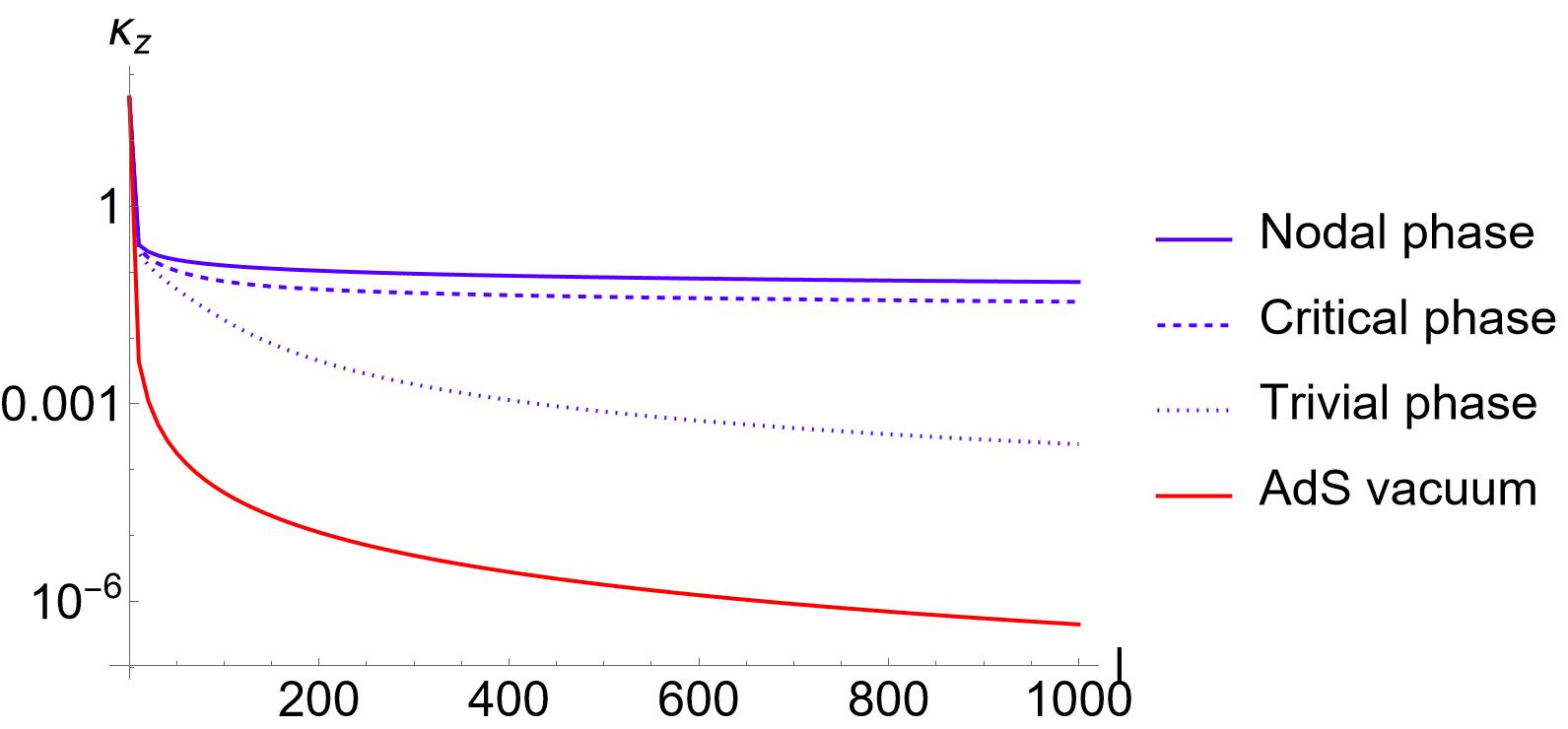}
            \end{minipage}
            \caption{Left: The evolution for the subtracted multi-entropy $\kappa_x$ with $l_x$ for different phases. Right: The evolution for the subtracted multi-entropy $\kappa_z$ with $l_z$ for different phases with representative values of $M/b$ in each phase.
           }
            \label{Figurel-kappa}
        \end{figure}

        To calculate $\kappa$ and analyze its properties, especially the large distance behavior, we consider the following configuration. We partition the boundary field theory into three regions: $A,B$ and $C$. Here $A$ and $B$ are two adjacent strips of equal width $l$, and $C$ is the complement of $A\cup B$. We then investigate the multipartite entanglement measure $\kappa(A:B:C)$ among these three regions. Fig.\,\ref{Figurel-kappa} plots the values of $\kappa$ versus $l$ for each phase and for the pure AdS background. It can be seen that $\kappa$ decreases with growing $l$ for all phases, which is consistent with general physical expectations. In the process that the length $l$ of $A$ and $B$ increases to $\infty$, the genuine tripartite entanglement among $ABC$ detected by $\kappa$ becomes smaller and smaller, which indicates the short range nature of the tripartite entanglement among $ABC$. Genuine tripartite entanglement among $A$, $B$ and the complementary subsystem $C$ vanishes at $l\to \infty$, consistent with the previous result that the state is an SRE state. This also implies that the state $ABC$ at $l\to \infty$ is a triangle state, with no genuine tripartite entanglement among $A$, $B$ and $C$,  where the tripartite entanglement emerges from bipartite entanglement among subsystems of $ABC$.

        Moreover, in the topologically non-trivial phase, $\kappa$ for strips in the $x$-direction is smaller than that of the vacuum state, while $\kappa$ for strips in the $z$-direction is larger. This occurs because the long range entanglement along the $xy$-directions is suppressed in the topological state, leading to a tripartite entanglement (among $A$, $B$, and their complement of $C$) that is weaker than that in the vacuum. In contrast, the $z$-direction retains a long scale correlation channel, so the tripartite multi-entropy there is actually larger than in the vacuum state.
        
\begin{figure}[htbp]
            \begin{minipage}{0.49\linewidth}
                \centering
                \includegraphics[width=\textwidth]{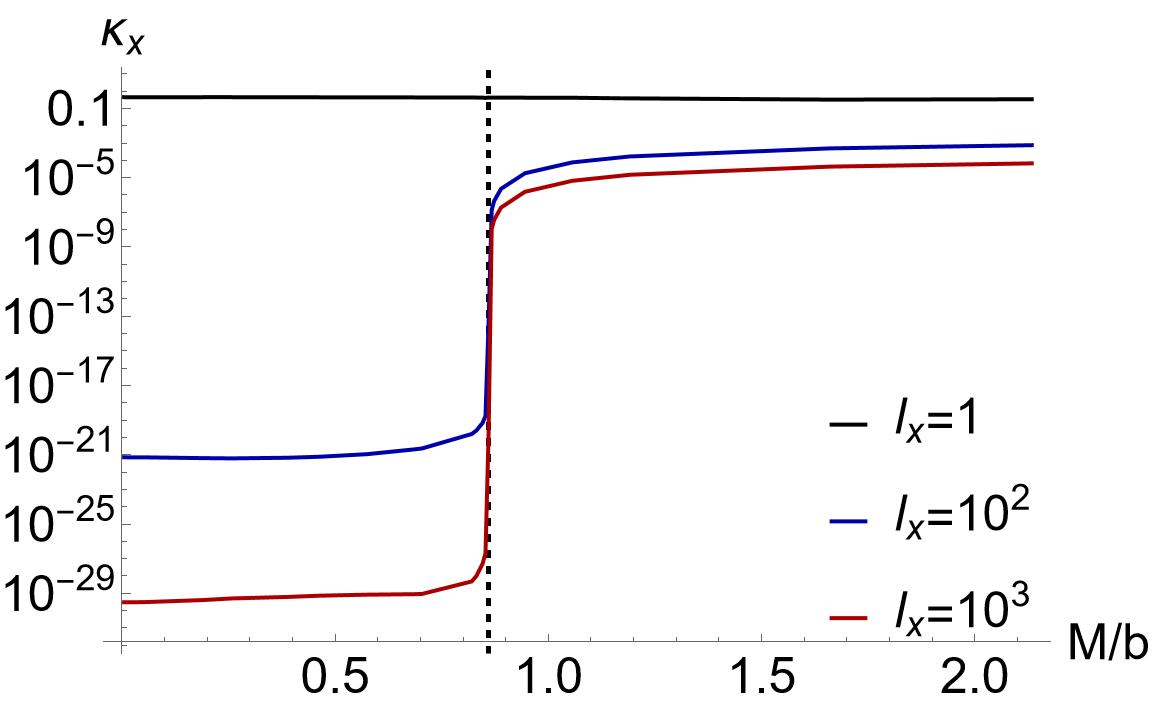}
            \end{minipage}
            \begin{minipage}{0.49\linewidth}
                \centering
                \includegraphics[width=\textwidth]{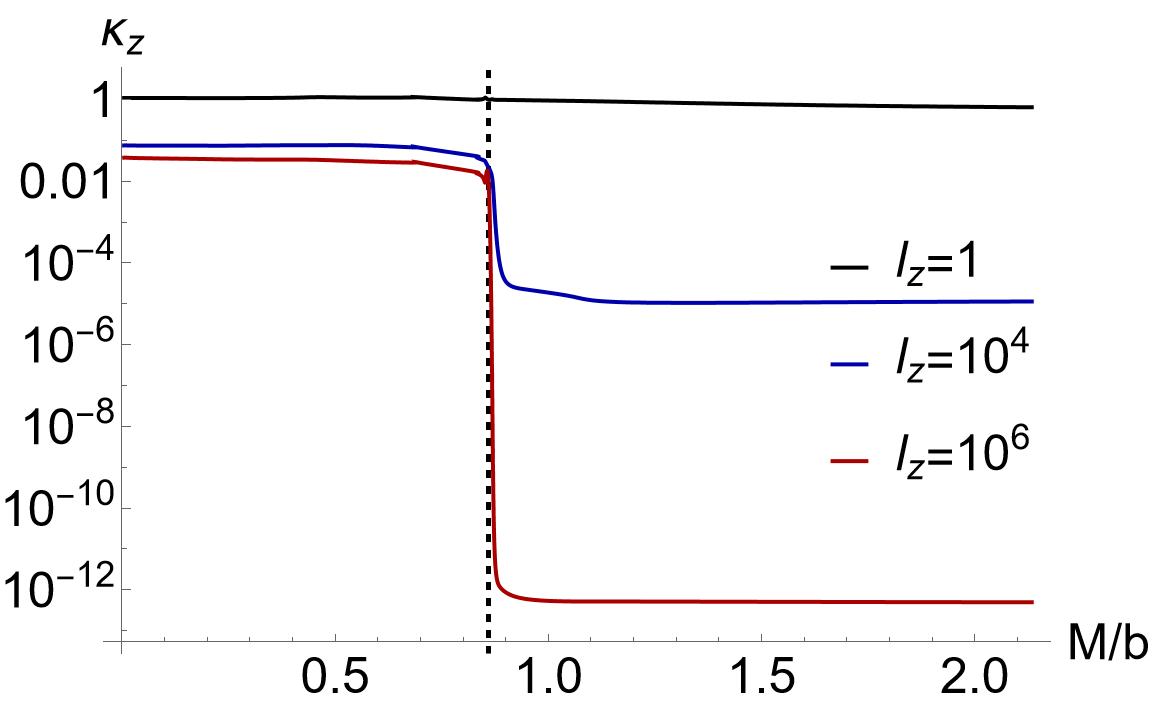}
            \end{minipage}
            \caption{Left: The evolution for $\kappa(A:B:C)$ with increasing $M/b$, where $A$ and $B$ are two strips aligned along $x$ direction with the same width and $C$ is their complement $C=(A\cup B)^c$. Right: The evolution for $\kappa(A:B:C)$ with increasing $M/b$, where $A$ and $B$ are two strips aligned along $z$ direction with the same width and $C$ is their complement $C=(A\cup B)^c$.}
            \label{FigureKappa}
        \end{figure}

        These features for kappa can be attributed to its anisotropic power-law behavior for large $l$, which is determined by the scalar behavior for corresponding IR geometry. When the strips $A$ and $B$ are aligned along the $x$-direction, the corresponding $\kappa$ scales as $l_x^{-1-\mathbf{z}}$; when the strips $A$ and $B$ are aligned along the $z$-direction, the corresponding $\kappa$ scales as $l_z^{-\frac{2}{\mathbf{z}}}$, where $\mathbf{z}$ takes the values 10.929, 6.3694, 1, and 1 as defined in \eqref{InfraRedScale} for the topologically nontrivial phase, the critical phase, the topologically trivial phase, and the AdS vacuum respectively. This explains the distinct behaviors of $\kappa$ among the phases observed in Fig.\,\ref{Figurel-kappa}. Moreover, because the scaling behavior differs from one phase to another, a sharp transition at the critical point naturally appears as shown in Fig.\,\ref{FigureKappa}.

        The values of $\kappa$ for several values of $l_x$ and $l_z$ as  a function of $M/b$ are shown in Fig.\,\ref{FigureKappa}. when the width of the strip region $A,B$ is small, the tripartite multi-entropy $\kappa$—after subtracting the vacuum contribution—does not vary with $M/b$. This trivial behavior can be explained by the fact that the corresponding Steiner tree fails to probe deep into the IR region. Physically, it reflects that when the $A,B$ strip is smaller than the correlation length, entanglement can penetrate directly, resulting in a trivial overall entanglement structure. As the width of the strip region $A,B$ increases, $\kappa$ decreases for every phase and exhibits a sharp transition at the critical point as $M/b$ is varied. For large enough $l$, $\kappa$ exhibits a sharp transition at the critical point in both the $x$ and $z$ directions. This indicates that the behavior of the genuine tripartite entanglement structure faithfully changes in the phase transition progress for the holographic nodal line semimetal. Moreover, the anisotropy behavior between $\kappa_x$ and $\kappa_z$ is the same with that of the CMI and the c-function.

\section{Multipartite entanglement from EWCS and the scaling behavior}
    \label{Section5}
    In this section, we introduce another class of multipartite entanglement measures based on the entanglement wedge cross section. In holographic duality, the EWCS ($E_W$) provides a geometric characterization of correlations that extends beyond the standard Ryu-Takayanagi surface, serving as the dual to both the reflected entropy $S_R(A:B)$ \cite{ReflectedEntropydutta2019canonicalpurificationentanglementwedge} and the entanglement of purification $E_P(A:B)$ \cite{Umemoto2018}. These quantities are specifically designed to isolate quantum entanglement in mixed-state bipartite states. The behavior of EWCS itself and various combinations constructed from the EWCS could be utilized as a detection of the underlying multipartite entanglement structures \cite{Bao2019,AnotherBao2019,AAnotherBao2019,ju2025entanglementwedgecrosssection,bao2026tripartitecorrelationsignalmultipartite}.
    We primarily focus on the EWCS as a probe of the IR physics in holographic nodal line semimetals, particularly analyzing its scaling behavior in the large-$l$ limit. 
    
    Furthermore, we utilize the EWCS to investigate a tripartite measure: the Markov Gap, defined as $h = S_R(A:B) - I(A:B)$. The Markov gap for a tripartite pure state is zero if the dual state is a sum of triangle states (SOTS) up to local unitary transformations \cite{SumOfTrangleStateZou2021,MarkovGapju2025holographicmultipartiteentanglementstructures}, so it quantifies  genuine tripartite entanglement that cannot be unitarily transformed to SOTS states. We will show that the large $l$ values of both the EWCS and the Markov gap exhibit sharp transitions at the quantum critical point in the holographic nodal line semimetal and could be used as probes for the quantum topological phase transition.
    
    \subsection{EWCS and its large \texorpdfstring{$l$}{\textit{l}} scaling behavior}
        In the holographic framework, the entanglement wedge cross-section is a significant geometric object. Simply put, given two boundary regions $A$ and $B$, within their entanglement wedge in the bulk, there always exists a cross-section that separates $A$ and $B$. The area of the minimal such cross-section $\gamma_{A,B}$ defines the EWCS, denoted as 
        \begin{equation}
            E_W(A:B)=\frac{1}{4G}{\rm Area}(\gamma_{A,B}),
        \end{equation}
        where $G$ is the gravitional constant. This geometric measure is important because it is dual both to the entanglement of purification $E_P(A:B)$ and to the reflected entropy $S_R(A:B)$ in the boundary field theory.
        \begin{equation}
            \label{HolographicReflectedEntropy}
            S_R(A:B)=2E_P(A:B)=2E_W(A:B).
        \end{equation}
        
\begin{figure}[htbp]
            \begin{minipage}{0.49\linewidth}
                \centering
                \includegraphics[width=\textwidth]{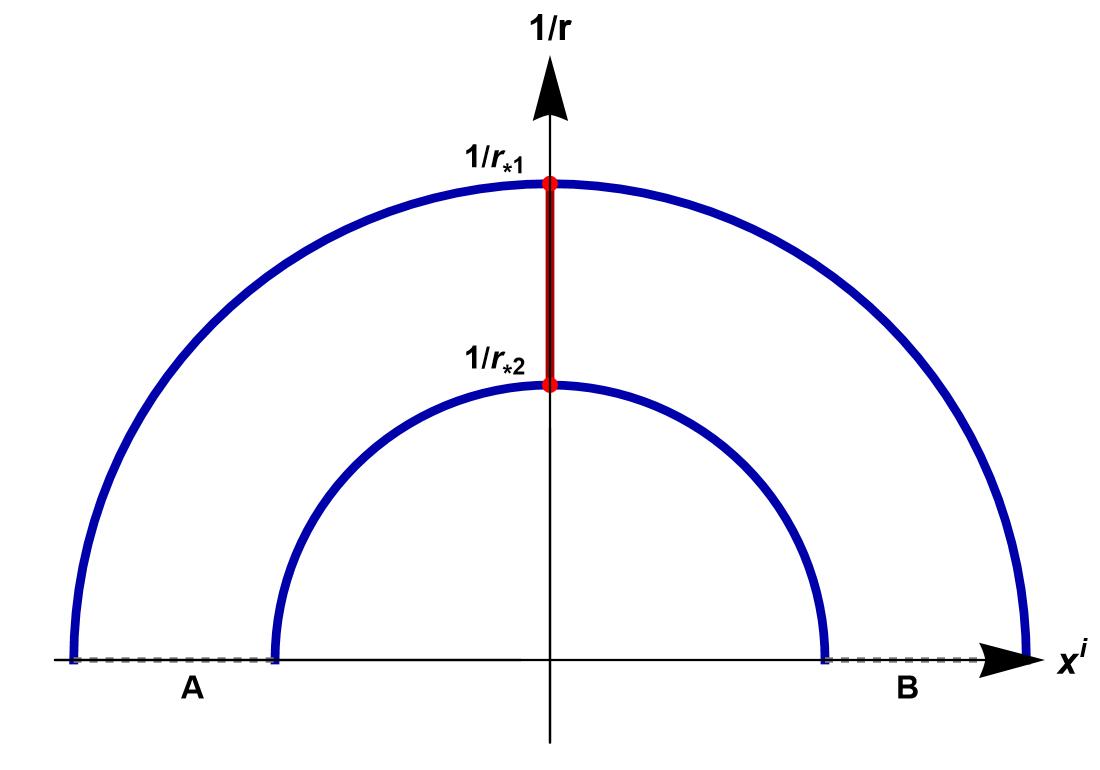}
            \end{minipage}
            \begin{minipage}{0.49\linewidth}
                \centering
                \includegraphics[width=\textwidth]{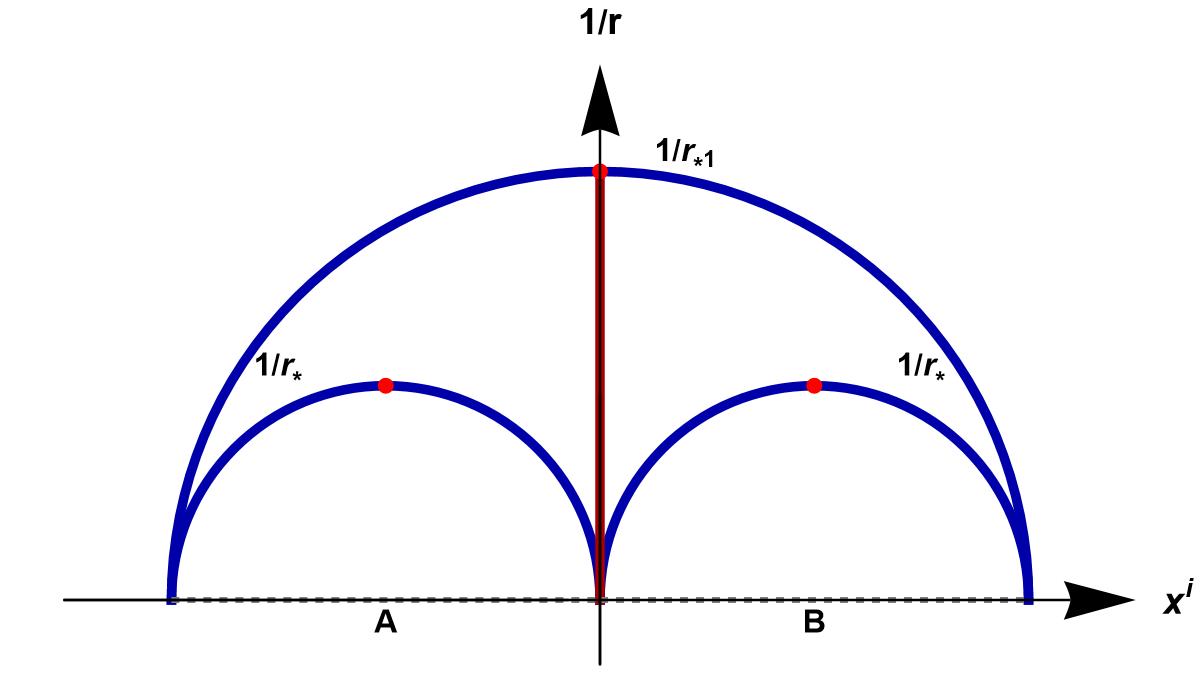}
            \end{minipage}
            \caption{Left: the RT surface for two disjoint strips $A$ and $B$ with the same width aligned along the $x^i$ direction (marked in blue) and the corresponding holographic EWCS (highlighted in red). Right: the RT surface for two adjacent strips $A$ and $B$ with the same width aligned along the $x^i$ direction (marked in blue) and the corresponding holographic EWCS (highlighted in red).}
            \label{HolographicEntanglementWedgeCrossSection}
        \end{figure}

        We now compute the entanglement wedge cross-section for two parallel strips, $A$ and $B$, of equal width aligned along the $x^i$-direction in the holographic nodal line semimetal. We consider the strips to be disjoint but sufficiently close such that their entanglement wedge is connected (i.e., the density matrix $\rho_{A\cup B}$ is not separable), as shown in the left panel of Fig.\,\ref{HolographicEntanglementWedgeCrossSection}. Due to the equal widths of the strips and the translational symmetry of the setup, the connected entanglement wedge possesses a reflection symmetry. This symmetry implies that the EWCS lies precisely on the wedge's mirror plane. Consequently, the protocol for calculating the EWCS in this symmetric configuration is as follows.
        \begin{enumerate}
            \item Select two turning points $r_{*1}$ and $r_{*2}$ sufficiently close to the horizon to capture the IR geometry and compute corresponding boundary width $l_1$ and $l_2$ via \ref{StripWidth}. The strip width for each $A$ or $B$ is then $(l_2-l_1)/2$.
            \item Adjust the value for $r_{*1}$ and $r_{*2}$ iteratively until the derived strip width for region $A$ and $B$ matches the desired scale.
            \item The area for the minimal entanglement wedge cross section can be compluted as 
                \begin{equation}
                    E_W(A:B)=\int_{r_{*1}}^{r_{*2}}\sqrt{g_{rr}\prod_{j\ne i}g_{jj}}dr.
                \end{equation}
        \end{enumerate}
         
        The case of two adjacent strips (Fig.\,\ref{HolographicEntanglementWedgeCrossSection}, right panel) corresponds to the limit where the intermediate turning point $r_{*2}$ approaches the asymptotic boundary.

        \begin{figure}[htbp]
            \begin{minipage}{0.49\linewidth}
                \centering
                \includegraphics[width=\textwidth]{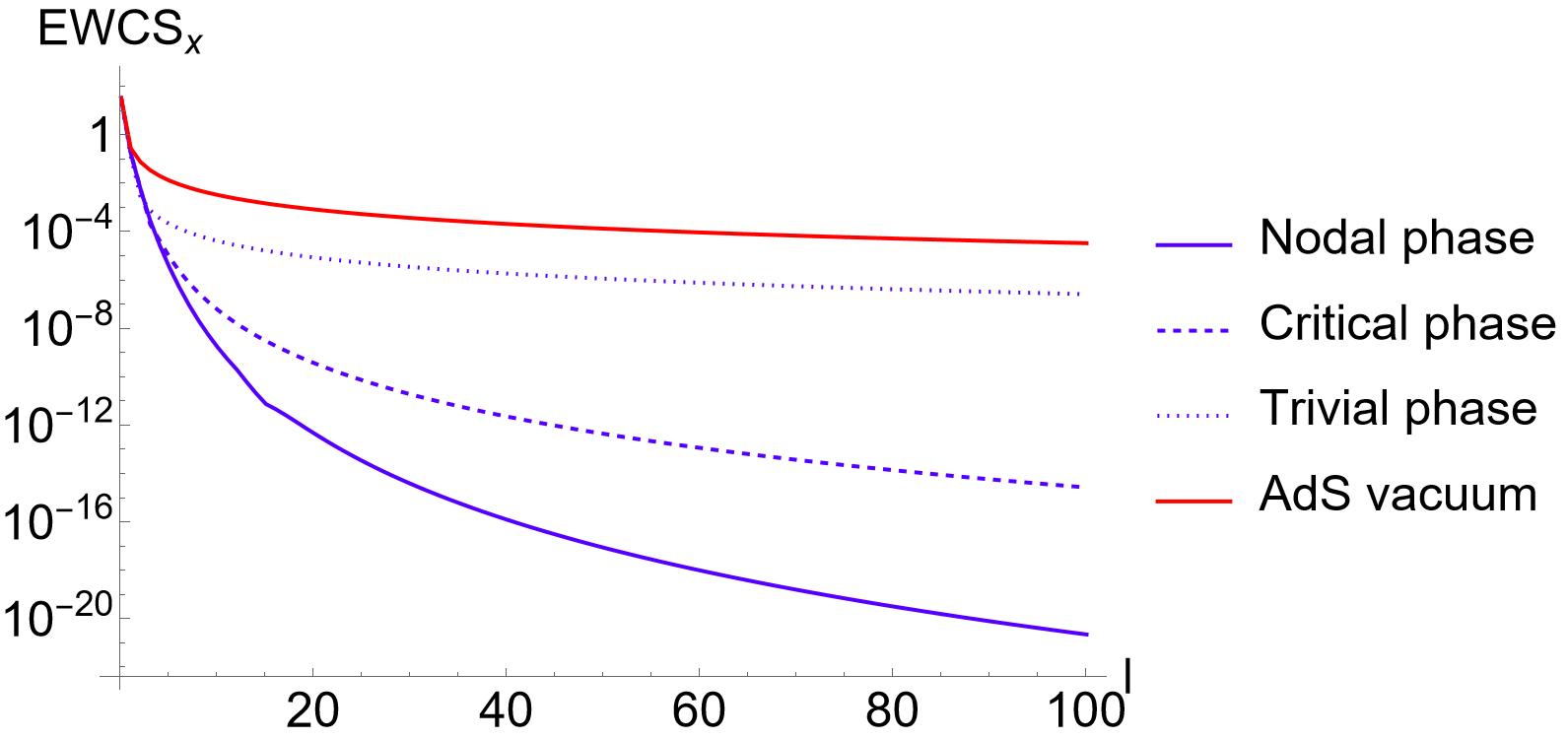}
            \end{minipage}
            \begin{minipage}{0.49\linewidth}
                \centering
                \includegraphics[width=\textwidth]{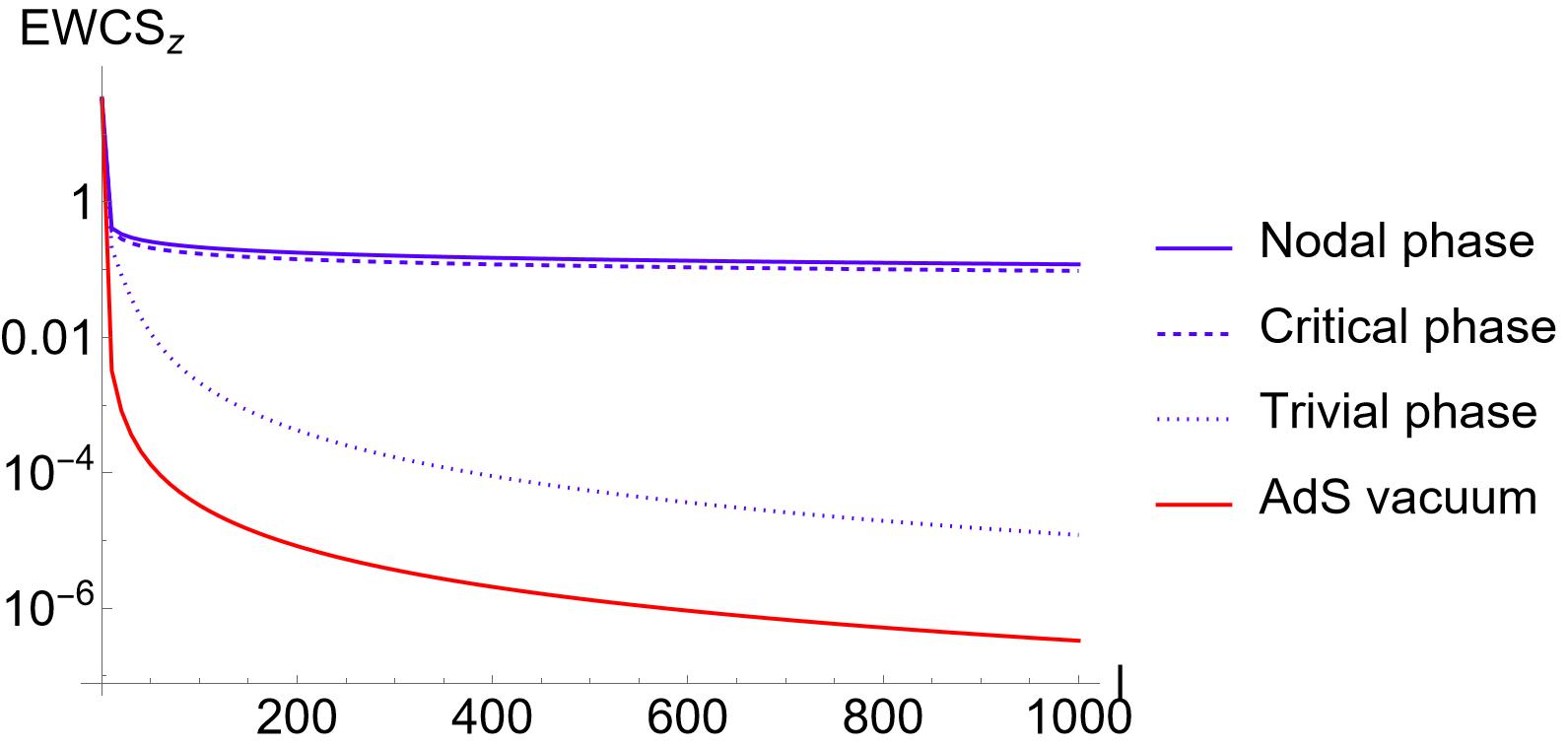}
            \end{minipage}
            \caption{Left: The evolution for the EWCS with $l_x$ for different phases. Right: The evolution for the EWCS with $l_z$ for different phases with representative values of $M/b$ in each phase.}
            \label{Figurel-EWCS}
        \end{figure}
        {For the configuration where the two strips are adjacent, the EWCS is UV‑divergent. Therefore, we consider instead a configuration with two non‑adjacent strips {with lengths $l$ at a  properly chosen distance, whose explicit value does not affect the qualitative behavior of the results,} and compute the dependence of the EWCS on $l$ for each phase, as shown in Fig.\,\ref{Figurel-EWCS}. It can be seen that when $l$ is large, the EWCS roughly follows a power-law in $l$. By fitting we obtain the leading power exponents: in the $x$-direction it behaves as $l_x^{-1-\mathbf{z}}$ , while in the $z$-direction it behaves as $l_z^{-\frac{2}{\mathbf{z}}}$, where $\mathbf{z}$ denotes the scaling exponent of the $z$-direction in the IR geometry. The values of $z$ for the topologically nontrivial, critical, and topologically trivial phases are $\mathbf{z}=\{\frac{2}{\alpha},~\frac{2}{\alpha_c},1\}=\{10.929,~ 6.36943,~ 1\}$, respectively, where $\alpha$ is defined in \eqref{NonTrivialInfraredGeometry} and $\alpha_c$ is defined in \eqref{CriticalInfraredGeometry}. This indicates that, owing to the anisotropy of the system, the entanglement in the topologically non‑trivial phase is strongly compressed in the $x$‑direction with a narrowed “throat”, whereas it remains open in the z‑direction. These different behaviors are precisely reflected in the scaling of the EWCS along the two directions.}
        \begin{figure}[htbp]
            \begin{minipage}{0.49\linewidth}
                \centering
                \includegraphics[width=\textwidth]{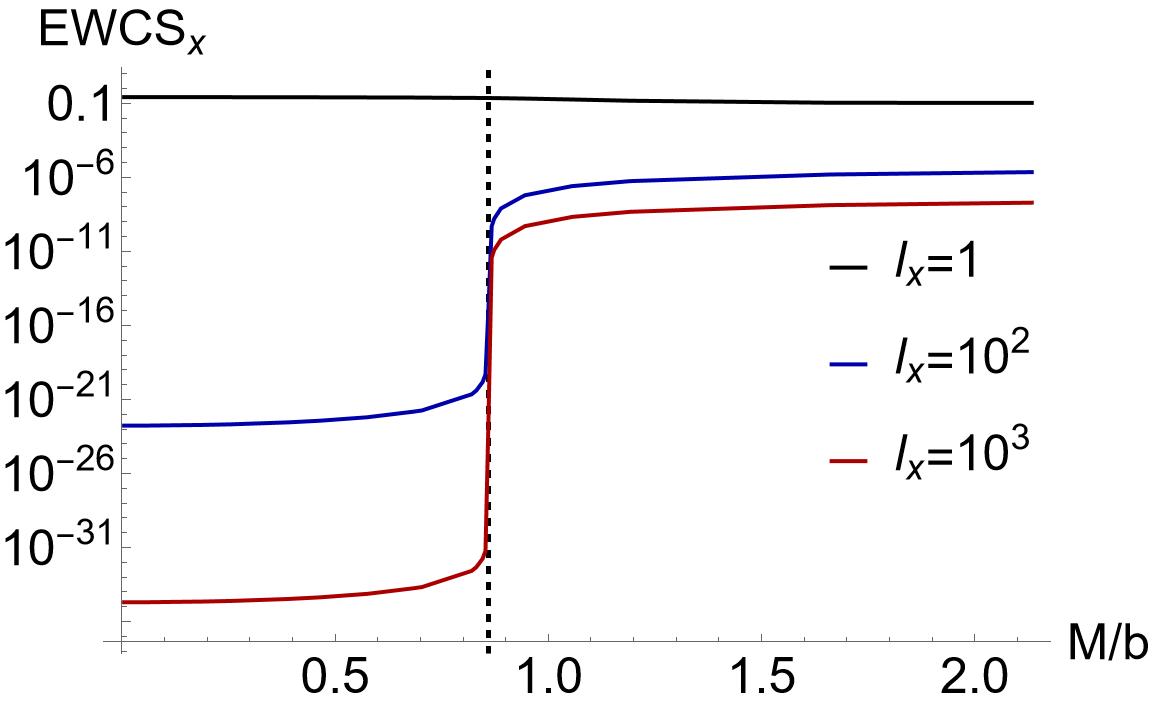}
            \end{minipage}
            \begin{minipage}{0.49\linewidth}
                \centering
                \includegraphics[width=\textwidth]{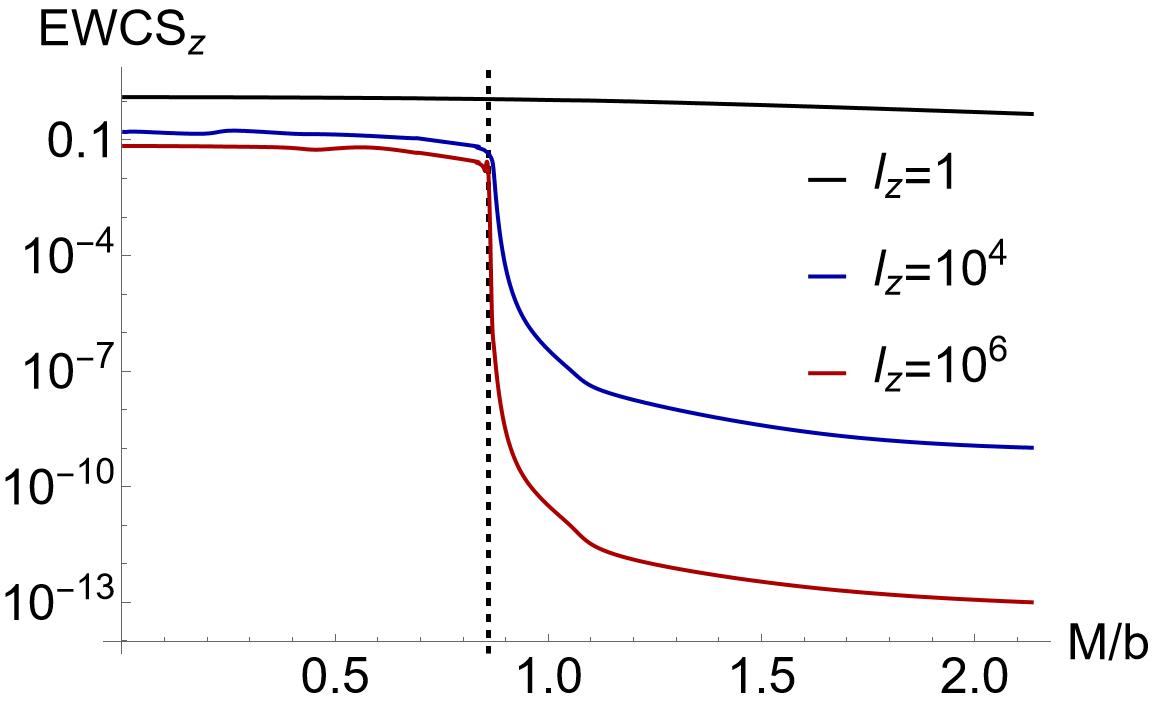}
            \end{minipage}
            \caption{The dependence of the EWCS $E_W(A:B)$ on values of $M/b$, where $A$ and $B$ are disadjacent strips aligned along the $x$-direction (left) and the $z$-direction (right) with the same width $l_x$ and $l_z$. }
            \label{FigureEWCS}
        \end{figure}
        
        We have also calculated the value of the EWCS as a function of $M/b$, which is shown in Fig.\,\ref{FigureEWCS}. When $l$ is small, the EWCS hardly varies with $M/b$. This is because the entanglement wedge at such scales does not probe deep enough into the IR region to detect differences between the phases. As $l$ increases, the EWCS begins to show distinct behavior among the phases. Specifically, in the $x$-direction, the EWCS grows with increasing $M/b$, whereas in the $z$-direction it decreases. This indicates that along the $x$-direction, as the system evolves from a topologically non-trivial phase to a topologically trivial one, the throat of the entanglement wedge gradually opens up and approaches the vacuum configuration. Conversely, along the $z$-direction, the throat of the entanglement wedge progressively narrows down as the system transitions from a topologically non-trivial to a trivial phase, eventually tending toward the vacuum case. 
        
        These observations are consistent with our earlier discussion on the scaling behavior of the EWCS. Moreover, they align with the conclusion drawn from previous entanglement measures-namely, that in the holographic nodal line semimetal, long range entanglement is suppressed in the $x/y$-directions, while a long range entanglement channel is preserved along the $z$-direction. These results could be interprested as the behavior of the reflected entropy or the entanglement of purification between two nonadjacent strips with width $l$ at the boundary. As the figure shows a sharp transition at the critical point, the values of EWCS at large values of $l$ could also be utilized as an order parameter for this quantum topological phase transition.
        
    \subsection{The Markov gap}
        This subsection introduces the Markov gap, a quantity rooted in the quantum Markov chain condition. We first define its generalization as a candidate measure for multipartite entanglement. We then compute its evolution for three adjacent parallel strips along the $x^i$-direction as a function of  $M/b$. In \cite{MarkovGapHayden_2021}, the Markov gap is defined as the difference between the reflected entropy and the mutual information
        \begin{equation}
            \label{MarkovGap}
            h(A:B)=S_R(A:B)-I(A:B),
        \end{equation}
        where $S_R(A:B)=2E_W(A:B)$ and $I(A:B)=S_A+S_B-S_{A\cup B}$. From the lower bound $S_R(A:B)\ge I(A:B)$, it follows that the Markov gap is non-negative, $h(A:B)\ge 0$.  As in holography, both $S_R(A:B)$ and $E_P(A:B)$ correspond to the same quantity: the EWCS of $A$ and $B$, the value of the Markov gap $h(A:B)$ is equivalent to another function \begin{equation}
         g(A:B)=2 E_P(A:B)-I(A:B)   \end{equation} in holography. Therefore, the entanglement structure that the Markov gap could detect could be explained from the properties of both these two functions. 
         
        It has been proved in \cite{SumOfTrangleStateZou2021} that $g(A:B)=0$ if and only if the quantum state is a triangle state \eqref{DefinitionTriangleState} up to local unitary transformations, while $h(A:B)=0$ if and only if the quantum state is a sum of triangle states (SOTS) \cite{SumOfTrangleStateZou2021} up to local unitary transformations. An SOTS state takes the following form
        \begin{equation}
            \label{SOTS}
            \ket{\psi}_{ABC}=\sum_{j}\sqrt{p_j}\ket{\psi_j}_{A^j_LB^j_R}\ket{\psi_j}_{B^j_LC^j_R}\ket{\psi_j}_{A^j_RC^j_L},
        \end{equation}
        where $\sum_jp_j=1$. A triangle state is a special case of an SOTS state. Therefore, the condition $g(A:B)=0$ is stronger than the condition $h(A:B)=0$ and $g(A:B)\geq h(A:B)\geq0$. For instance, the tripartite GHZ state has vanishing $h(A:B)$ but non-vanishing $g(A:B)$, and the GHZ state is indeed an SOTS state but not a triangle state. On the other hand, the condition $h(A:B)\neq 0$ is stronger than the condition $g(A:B)\neq 0$. When $h(A:B) \neq 0$, the system should have tripartite entanglement structures that are neither triangle states nor SOTS states, while when $g(A:B)\neq 0$, there could still exist non-triangle SOTS entanglement structures.

        As both of the two functions have the same holographic dual $2E_W(A:B)-I(A:B)$, in the holographic context, we should always adopt the stronger conditions in the two cases $2E_W(A:B)-I(A:B)=0$ and $2E_W(A:B)-I(A:B)\neq 0$. Therefore, when the Markov gap is not zero, it means that there are non-SOTS entanglement structures in the holographic system, while when the Markov gap is zero, the system should belong to a triangle state. Thus, the Markov gap severs as a measure of genuine tripartite entanglement in the holographic system.

        \begin{figure}
            \centering
            \includegraphics[width=0.75\textwidth]{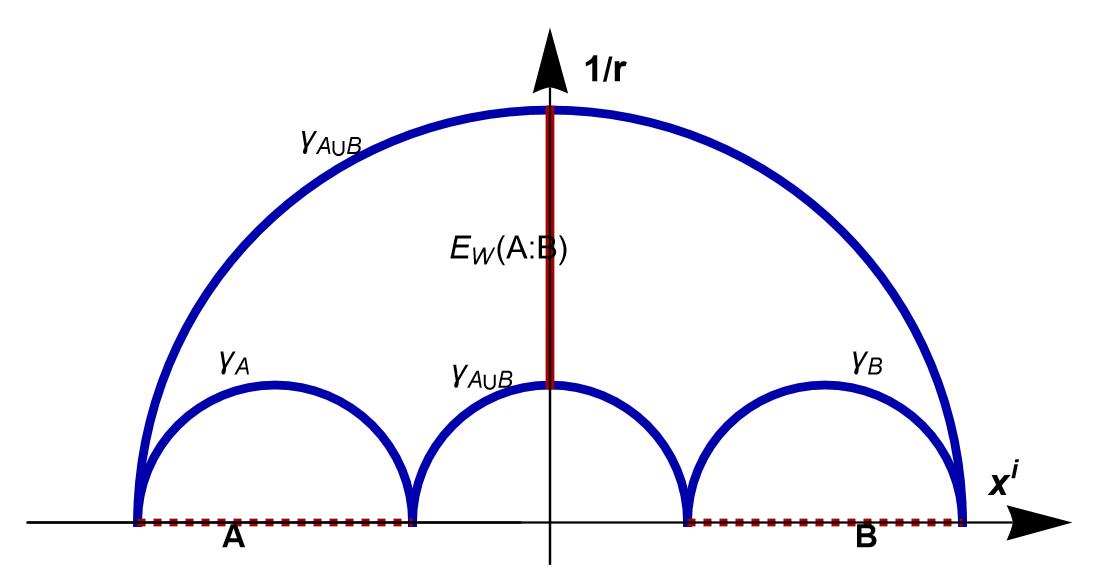}
            \caption{The Markov gap for two adjacent strips $A,B$ aligned along $x^i$-direction with the same width. The RT surface shown in red is the entanglement wedge cross section for $A$ and $B$, and the Markov gap for $A,B$ is computed as $2E_W(A:B)-({\rm Area\,}(\gamma_A)+{\rm Area\,}(\gamma_B)-{\rm Area\,}(\gamma_{A\cup B}))$.}
            \label{FigureMarkovGapConfiguration}
        \end{figure}

        Due to the property above, the Markov gap is expected to be a measure for tripartite entanglement among $A$, $B$ and $C$ of a pure state $ABC$. However, as a tripartite measure, we need a quantity that has permutation symmetry in $A$, $B$ and $C$. One choice is to build the geometric average value of $h(A:B)$, $h(B:C)$ and $h(C:A)$ as suggested in \cite{EWCSbasak2025newgenuinemultipartiteentanglement}. However, this is not a good choice here as for the configuration where one of $A$, $B$ and $C$ has to be a strip with infinite length, two terms in $h(A:B)$, $h(B:C)$ and $h(C:A)$ would be divergent and this contradicts the physical condition that tripartite entanglement between two subregions with lengths $l$ and their complement should vanish at $l\to \infty$. Therefore, we employ the following more reasonable definition of the permutation symmetric version of Markov gap
        \begin{equation}
            \label{GeneralizedMarkovGap}
            h(A:B:C)=\min\{h(A:B),~h(B:C),~h(C:A)\},
        \end{equation} which is consistent with the requirement above.
       
    \subsection{The large \texorpdfstring{$l$}{\textit{l}} scaling behavior as a probe of topological phase transitions}  
        \begin{figure}[htbp]
            \begin{minipage}{0.49\linewidth}
                \centering
                \includegraphics[width=\textwidth]{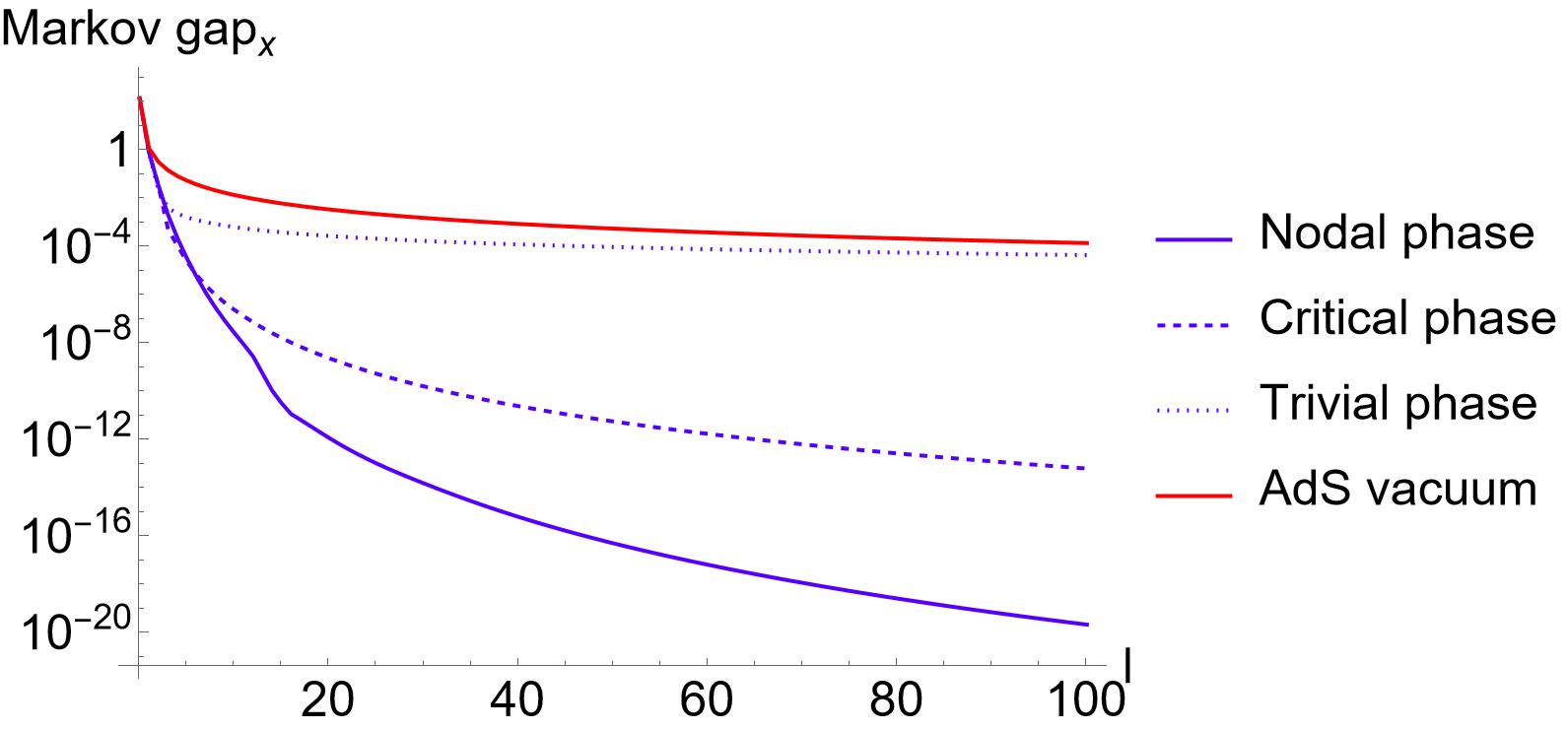}
            \end{minipage}
            \begin{minipage}{0.49\linewidth}
                \centering
                \includegraphics[width=\textwidth]{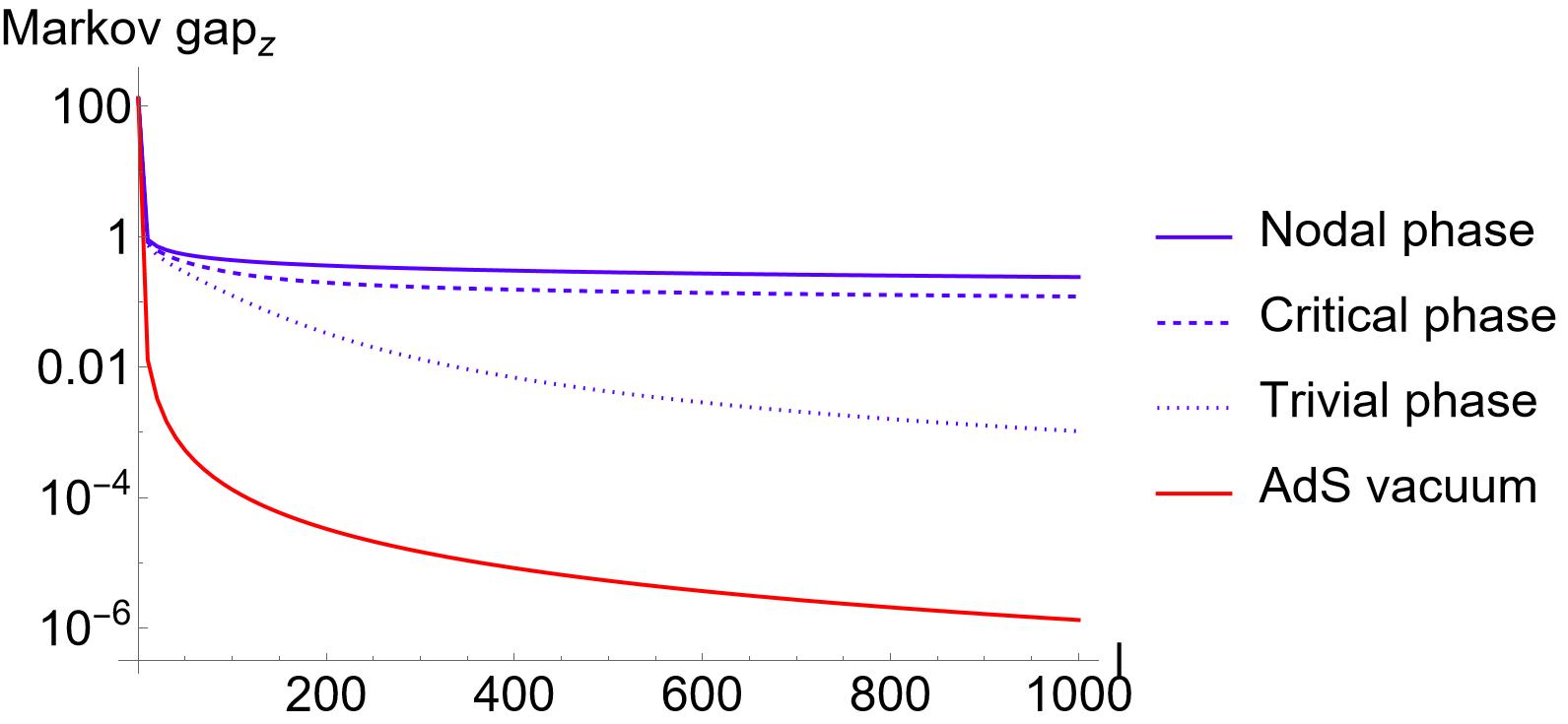}
            \end{minipage}
            \caption{left: The $l_x$ (left) and $l_z$ (right) dependence of the Markov gap $h(A:B:C)$ in $x$ and $z$ directions for different phases with representative values of $M/b$ in each phase.}
            \label{Figurel-MarkovGap}
        \end{figure}
       
        By choosing an appropriate configuration, we can greatly simplify the calculation of the  Markov gap. Here we select both $A$ and $B$ as strips of width $l$, with a separation also equal to $l$, and denote the complement of $A\cup B$ as the third party $C$, as illustrated in Fig.\,\ref{FigureMarkovGapConfiguration}. It is then straightforward to see that in this case $h(A:B:C)$ reduces to $h(A:B)$, allowing us to perform explicit computations.

        Fig.\,\ref{Figurel-MarkovGap} displays the variation of the Markov gap with the scale $l_x$ and $l_z$ for each phase and for the pure AdS background in both the $x$ and $z$ directions. At very small $l$ the curves of all phases coincide with the vacuum curve, consistent with our earlier diagnosis. As $l$ grows, we find that for a strip oriented along the $x$-direction the value in the topologically non-trivial phase lies below the vacuum value, and as the system transitions to the topologically trivial case it gradually approaches the vacuum curve. For a strip along the $z$-direction the trend is reversed.
        
        This occurs because of the anisotropy of the system. When the strip is along the $x$-direction, the IR behavior of $f(r)$ strongly pinches the “throat” of the entanglement wedge corresponding to $A\cup B$, resulting in a Markov gap smaller than that of the vacuum phase. In the $z$-direction, the infrared behavior of $u(r)$ keeps the throat of the $A\cup B$ entanglement wedge open and even expanding, so the Markov gap becomes larger than in the vacuum phase. This observation once again confirms that in the holographic nodal line system, long range entanglement along the $xy$-directions is suppressed, while becomes enhanced along the $z$-direction.
       
        The observed anisotropy of the Markov gap can be more precisely interpreted through its corresponding scaling behavior. It can be seen that when $l$ is large, the Markov gap roughly follows a power-law in $l$. By fitting we obtain the leading power exponents: in the $x$-direction it behaves as $l_x^{-1-\mathbf{z}}$ , while in the $z$-direction it behaves as $l_z^{-\frac{2}{\mathbf{z}}}$, where $\mathbf{z}$ denotes the scaling exponent of the $z$-direction in the IR geometry. The values of $z$ for the topologically nontrivial, critical, and topologically trivial phases are $\mathbf{z}=\{10.929,~ 6.36943,~ 1\}$, respectively.
        \begin{figure}[htbp]
            \begin{minipage}{0.49\linewidth}
                \centering
                \includegraphics[width=\textwidth]{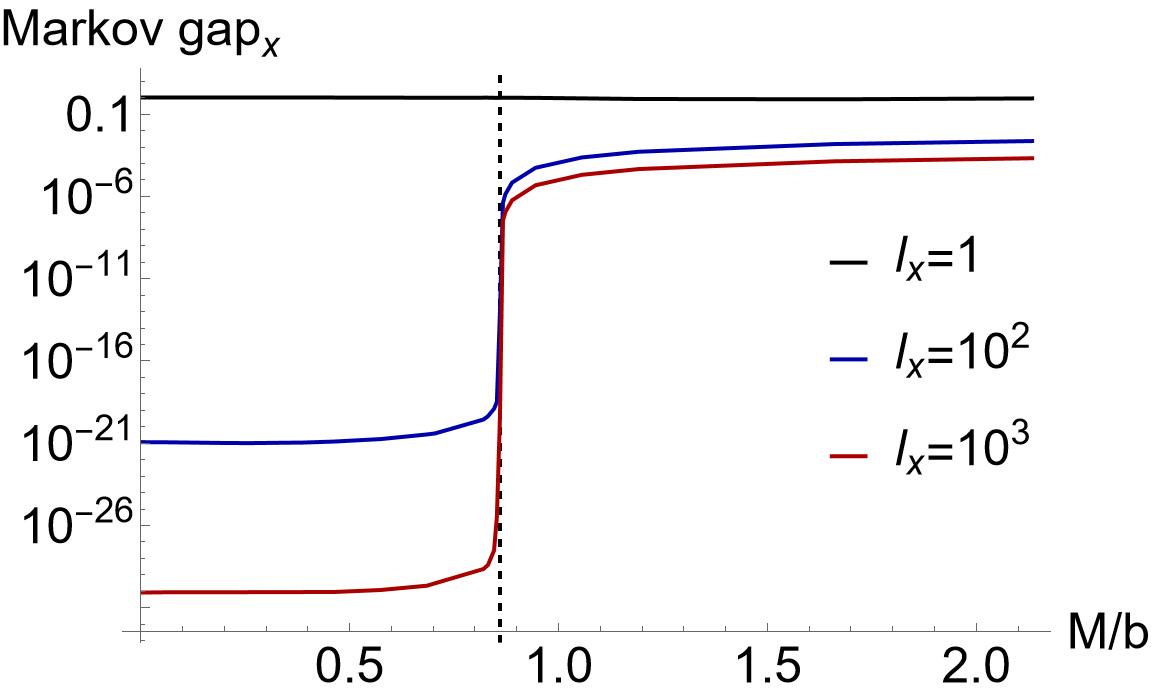}
            \end{minipage}
            \begin{minipage}{0.49\linewidth}
                \centering
                \includegraphics[width=\textwidth]{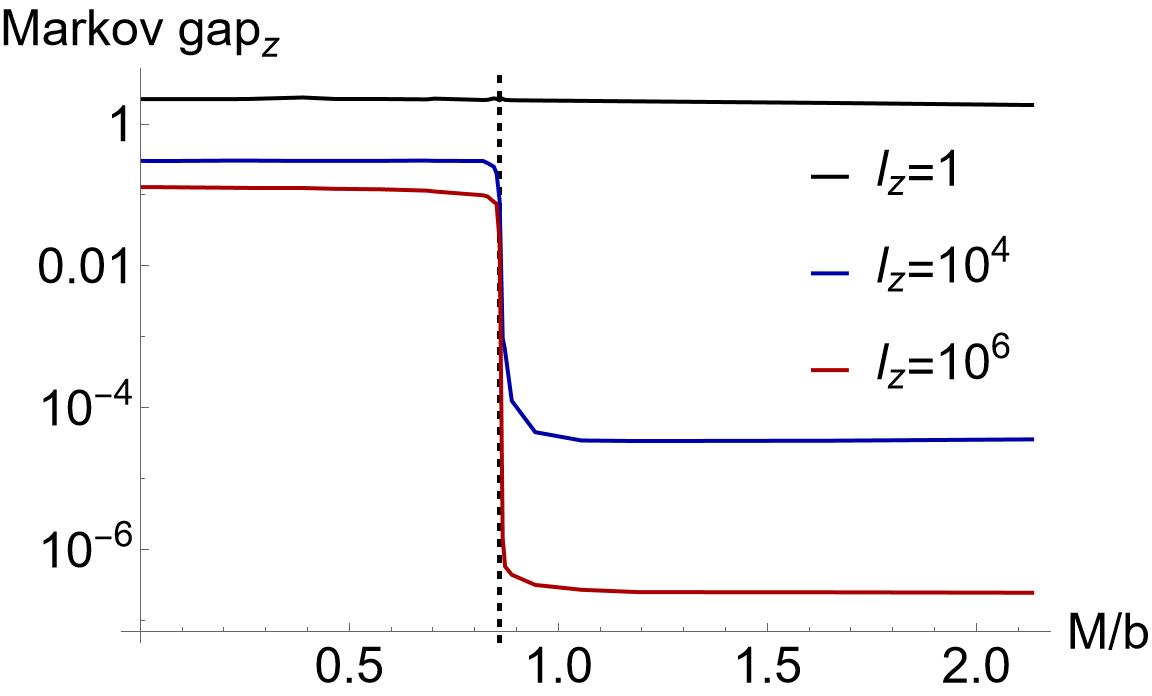}
            \end{minipage}
            \caption{Left: the evolution for the Markov gap $h(A,B,C)$ with increasing $M/b$ where $A,~B$ are strips aligned along the $x$ direction with the same width. Right: The evolution for the Markov gap $h(A,B,C)$ with increasing $M/b$ in the $z$ direction.}
            \label{FigureMarkovGap}
        \end{figure}
        
        The dependence of the Markov gap $h(A:B:C)$ on values of $M/b$ at fixed values of $l$ is shown in Fig.\,\ref{FigureMarkovGap}. At small $l$, the Markov gap does not vary with $M/b$. This is because the entanglement wedge does not yet penetrate deeply into the IR region and only probes the geometry of the asymptotic AdS regime, which is essentially the same for all phases. As $l$ increases, the Markov gap begins to show differences among the phases and exhibits a sharp transition at the critical point. When the strips are aligned along the $x$-direction, the Markov gap is larger in the trivial phase than in the non-trivial phase, while the opposite holds when the strip is along the $z$-direction. This distinct behavior in different phases makes the Markov gap also a good candidate as the order parameter in the holographic nodal line semimetal system.

\section{Discussions and outlook}
    \label{Section6}
    In this work, we have systematically investigated the entanglement structure of holographic nodal line semimetals, with a particular emphasis on multi-partite entanglement. By utilizing a diverse set of entanglement measures including the conditional mutual information, $\kappa$ constructed from the multi-entropy and the EWCS along with its associated Markov gap, we have analyzed the tripartite entanglement structures of the system across the topological phase diagram. 
    \begin{table}[ht]
        \centering
        \begin{tabular}{c c c c c c}
            \toprule
                & \textbf{c function} & \textbf{CMI} & $\boldsymbol{\kappa}$ & \textbf{EWCS} & \textbf{Markov gap} \\
            \midrule
                $x$-direction & $l_x^{1-\mathbf{z}}$ & $l_x^{-3-\mathbf{z}}$ &  $l_x^{-1-\mathbf{z}}$ & $l_x^{-1-\mathbf{z}}$ & $l_x^{-1-\mathbf{z}}$\\
                $z$-direction & $l_z^{2-\frac{2}{\mathbf{z}}}$ & $l_z^{-2-\frac{2}{\mathbf{z}}}$ & $l_z^{-\frac{2}{\mathbf{z}}}$ & $l_x^{-\frac{2}{\mathbf{z}}}$ & $l_z^{-\frac{2}{\mathbf{z}}}$\\
            \bottomrule
        \end{tabular}
        \caption{The scaling behaviours for each entanglement measures. Here $\mathbf{z}$ denotes the ratio of the leading-order exponents of $u(r)$ and $f(r)$ in the infrared geometry; its values for the topologically nontrivial, critical, and topologically trivial phases are $\mathbf{z}=\{\frac{2}{\alpha},~\frac{2}{\alpha_c},1\}=\{10.929,~ 6.36943,~ 1\}$, respectively, where $\alpha$ is defined in \eqref{NonTrivialInfraredGeometry} and $\alpha_c$ is defined in \eqref{CriticalInfraredGeometry}.}
        \label{ScalingBehaviourTable}
    \end{table}
    
    Our results confirm that the holographic nodal line semimetal, despite its strong coupling, remains a symmetry-protected short-range entangled state rather than a long-range entangled state, as evidenced by the vanishing of all considered entanglement measures in the long-distance ($l \to \infty$) limit. However, we demonstrate that the asymptotic scaling behavior of these measures at large $l$ serves as a definitive signature of the IR physics. While the absolute entanglement values vanish, the power law exponents and the rate of decay encode the presence of the topological nodal structures. These scaling laws, summarized in Table \ref{ScalingBehaviourTable}, undergo sharp transitions at the critical point, establishing multipartite entanglement measures as robust non-local order parameters for quantum topological phase transitions beyond the Landau paradigm.

    Furthermore, our analysis reveals a significant spatial anisotropy in the entanglement structure. For a nodal ring lying in the $k_x-k_y$ plane, correlations in the $x$-$y$ directions are suppressed by the freezing of degrees of freedom in the IR, whereas the entanglement in the $z$-direction gets enhanced. 
    
    There are several open questions. First, a natural extension is to apply this multipartite entanglement framework to holographic Weyl semimetals and more complex systems, such as the holographic Weyl-$Z_2$ semimetal \cite{Ji:2021aan} or the Weyl-Nodal line coexisting semimetals \cite{HolographicWeylNodalChu_2024}. These systems possess richer phase structures where multiple topological phases exist. Investigating their entanglement structure provides more understanding on the strongly coupled topological semimetal systems. Second, while this work focused on SRE phases, a fundamental question remains: what modifications to the IR geometry can induce a transition to a truly long-range entangled state dual to topological order\cite{naskar2024topologicalentanglemententropymeets}? Third, the behavior of multipartite entanglement during a quantum quench remains largely unexplored in holographic semimetals. Observing how the entanglement scaling behavior forms or collapses in real-time could offer new perspectives on the stability of topological features under thermalization.

\subsection*{Acknowledgments}
We thank Xin-Xiang Ju, Karl Landsteiner, Wen-Peng Li, Bo-Hao Liu, Yan Liu, Wen-Bin Pan, Xin-Meng Wu and Yang Zhao for helpful discussions. This work was supported by the National Natural Science Foundation of China (Grant Nos. 12035016, 12405078, 12575068).

\bibliographystyle{elsarticle-num}
\bibliography{biblio}

\end{document}